\documentclass[fleqn,usenatbib]{mnras}

\usepackage{pdflscape}

\usepackage[T1]{fontenc}

\DeclareRobustCommand{\VAN}[3]{#2}
\let\VANthebibliography\thebibliography
\def\thebibliography{\DeclareRobustCommand{\VAN}[3]{##3}\VANthebibliography}

\usepackage{graphicx}	
\usepackage{amsmath}	
\usepackage{amssymb}	
\usepackage{subfigure}  
\usepackage{threeparttable}
\usepackage{multirow}
\usepackage{inputenc}
\usepackage{xcolor}
\usepackage{adjustbox}
\usepackage[separate-uncertainty = true,multi-part-units=single,group-separator={,}, group-digits = integer, group-four-digits = true,per-mode=reciprocal]{siunitx}
\usepackage{chemformula}
\usepackage{float}

\usepackage{newtxtext,newtxmath}

\title[NIRISS and HST Comparison]{JWST/NIRISS and HST: Exploring the improved ability to characterise exoplanet atmospheres in the JWST era}

\author[]{Chloe Fisher$^{1}$\thanks{E-mail: chloe.fisher@physics.ox.ac.uk},
Jake Taylor$^{1}$\thanks{E-mail: jake.taylor@physics.ox.ac.uk}, 
Vivien Parmentier$^{2}$, 
Daniel Kitzmann$^{3,4}$, 
Jayne L. Birkby$^{1}$,
Michael Radica$^{5}$,
\newauthor
Joanna Barstow$^{6}$,
Jingxuan Yang$^{7}$,
Giuseppe Morello$^{8}$
\\
$^{1}$Astrophysics, University of Oxford, Denys Wilkinson Building, Keble Road, Oxford, OX1 3RH, United Kingdom\\
$^{2}$Universit\'e C\^ote d'Azur, Observatoire de la C\^ote d'Azur, CNRS, Laboratoire Lagrange, France\\
$^{3}$Space Research and Planetary Sciences, Physics Institute, University of Bern, Gesellschaftsstrasse 6, 3012 Bern, Switzerland\\
$^{4}$Center for Space and Habitability, University of Bern, Gesellschaftsstrasse 6, 3012 Bern, Switzerland\\
$^{5}$Trottier Institut de Recherche sur les Exoplanètes (iREx), Université de Montréal, 1375 Avenue Thérèse-Lavoie-Roux, Montréal,
QC, H2V 0B3, Canada\\
$^{6}$School of Physical Sciences, The Open University, Walton Hall, Milton Keynes, MK7 6AA, UK\\
$^{7}$Atmospheric, Oceanic and Planetary Physics, Department of Physics, University of Oxford, Oxford OX1 3PU, UK\\
$^{8}$Instituto de Astrof\'isica de Andaluc\'ia (IAA-CSIC), Glorieta de la Astronom\'ia s/n, 18008 Granada, Spain
}

\date{Accepted XXX. Received YYY; in original form ZZZ}

\pubyear{2024}

\begin{document}
\label{firstpage}
\pagerange{\pageref{firstpage}--\pageref{lastpage}}
\maketitle

\begin{abstract}
The Hubble Space Telescope has been a pioneering instrument for studying the atmospheres of exoplanets, specifically its WFC3 and STIS instruments. With the launch of JWST, we are able to observe larger spectral ranges at higher precision. NIRISS/SOSS covers the range 0.6--2.8 microns, and thus can serve as a direct comparison to WFC3 (0.8--1.7 microns). We perform atmospheric retrievals of WFC3 and NIRISS transmission spectra of WASP-39 b in order to compare their constraining power. We find that NIRISS is able to retrieve precise \ch{H2O} abundances that do not suffer a degeneracy with the continuum level, due to the coverage of multiple spectral features. We also combine these datasets with spectra from STIS, and find that challenges associated with fitting the steep optical slope can bias the retrieval results. In an effort to diagnose the differences between the WFC3 and NIRISS retrievals, we perform the analysis again on the NIRISS data cut to the same wavelength range as WFC3. We find that the water abundance is in strong disagreement with both the WFC3 and full NIRISS retrievals, highlighting the importance of wide wavelength coverage. Finally, we carry out mock retrievals on the different instruments, which shows further evidence of the challenges in constraining water abundance from the WFC3 data alone. Our study demonstrates the vast information gain of JWST's NIRISS instrument over WFC3, highlighting the insights to be obtained from our new era of space-based instruments.  
\end{abstract}




\section{Introduction}

The arrival of data from JWST presents a huge increase in the resolution and wavelength coverage of exoplanet atmosphere spectra we can obtain from space-based observatories. Spanning from \SI{0.6}{\um} all the way to \SI{28}{\um} across four different instruments with spectroscopic capabilities, the gain in potential information for exoplanet atmospheres is no doubt extensive. This new wavelength domain has facilitated the discovery of new molecules and confirmed previous detections \citep[e.g.,][]{bell23}, improved our analysis of cloudy objects \citep[e.g.,][]{kempton23,grant23}, and pushed the limits of the smallest planets we can observe \citep[e.g.,][]{lustig-yaeger23,moran23,kirk24}. In the previous few decades, our space-based spectroscopic analyses of exoplanet atmospheres have been limited to the Hubble (HST) and Spitzer Space Telescopes. One of the most powerful tools of characterisation with HST has been via transmission spectra. HST is able to provide transmission spectra in the 0.2--\SI{1.7}{\um} range with the Wide Field Camera 3 (WFC3) instrument, via its UVIS and NIR channels (0.2--\SI{1.0}{\um} and 0.85--\SI{1.7}{\um}, respectively). Its Space Telescope Imaging Spectrograph (STIS) can also measure transmission spectra, in the 0.115--\SI{1.0}{\um} range. 

Transmission spectra from WFC3 have been used considerably over the years to characterise the atmospheres of exoplanets, with the first observation being the super-Earth GJ 1214b \citep{berta12}, yielding a featureless spectrum. Over the last 10 years, the majority of the WFC3 spectral analysis has focused on measuring the water (or methane) feature at \SI{1.4}{\um} \citep[e.g.,][]{deming13,wakeford13,kreidberg14a}. However, various studies have also looked at the signatures from clouds and hazes in WFC3 transmission spectra \citep[e.g.,][]{gibson12,kreidberg14,sing15}. As the number of published transmission spectra from WFC3 grew, statistical studies emerged, comparing results across a range of exoplanets, to explore potential trends between parameters such as metallicity and mass, water abundance, clouds, and temperature \citep[e.g.][]{sing16,barstow17,macdonald17b,tsiaras18,fisher18,pinhas19,welbanks19}.

For a single JWST instrument, the closest comparison to WFC3 is the Near Infrared Imager and Slitless Spectrograph (NIRISS) \citep{doyon23}, covering 0.6--\SI{2.8}{\um} with its Single Object Slitless Spectroscopy (SOSS) mode \citep{albert23}, capable of providing transmission spectra. Note that although the G140 mode of the Near Infrared Spectrograph (NIRSpec) covers 0.7--\SI{1.84}{\um}, and could also be compared to WFC3, there are no exoplanet transmission observations using it, either current or planned, since NIRISS is a more optimal instrument for the majority of cases. Due to its two filters, G140 can only observe either 0.7--\SI{1.27}{\um} or 0.97--\SI{1.84}{\um}, so multiple transits would need to be taken to cover the full wavelength range. Nevertheless, NIRISS can face issues with contaminated fields or saturation with bright stars, so may not be suitable for all targets. However, newly available in the upcoming JWST cycle 4 is the Near Infrared Camera's (NIRCam) short wavelength grism time-series mode \citep{schlawin16}, allowing spectroscopic observations of exoplanets around bright stars at 0.6--\SI{2.3}{\um}. This could provide an ideal option for targets on which NIRISS would saturate.

Approximately 27 proposals for exoplanet transmission in JWST cycles 1, 2 and 3 will use NIRISS/SOSS, leading to many opportunities to take advantage of what this instrument has to offer. In light of previous WFC3 results and current and upcoming NIRISS observations, a comparative question about the instruments presents itself -- can we quantify the information gained from NIRISS compared to WFC3? And furthermore, can we determine the source of this information? These are the key questions motivating the current study.

The hot gas giant WASP-39 b was chosen as a case study for the JWST Transiting Exoplanet Early-Release Science (ERS) program, which provided the community with transmission spectra of this planet from several instruments, including NIRISS. This data, along with previous observations of WASP-39 b with HST \citep{wakeford18}, provides the opportunity to perform a direct comparison of NIRISS and WFC3 data. In this study, we compare constraints from atmospheric retrievals of the NIRISS and WFC3 transmission spectra of WASP-39 b. Furthermore, we attempt to determine the driving factor behind the information gain for each model parameter. Since the NIRISS wavelength range entirely encompasses that of WFC3, by cutting the NIRISS spectra down to this range, we can test whether the increase in wavelength coverage or spectral resolution is more important. We stress that the intention of this study is to compare the information content of WFC3 and NIRISS, rather than to make confident conclusions about the nature of the atmosphere of WASP-39 b.

Although JWST has a number of instruments spanning a wide wavelength range, it lacks coverage of the optical wavelengths. Several studies have shown the importance of optical data in characterising exoplanets through transmission spectroscopy \citep[e.g.][]{lecavelierdesetangs08,benneke12,griffith14}, and recently \cite{fairman24} demonstrated the complimentary nature of HST/STIS (covering the 0.2--\SI{1.0}{\um} range) when combined with JWST data, in particular for constraining cloud parameters and alkali species. However, combining spectra from different instruments can be challenging \citep[e.g.,][]{yip21}, particularly if the data are reduced in distinct ways assuming different orbital parameters for the planetary system. In this work, we test the effect of combining the NIRISS and WFC3 transmission spectra of WASP-39 b with data from STIS, to establish if this can add useful information in either case, and assess the complications that can arise from combining datasets with inconsistent reduction methods.

\subsection{Exoplanets in transmission with NIRISS/SOSS}

Since the start of JWST observations, NIRISS has been used in a number of exoplanet transmission studies, the first of which was obtained through the Early Release Observations (ERO) of HAT-P-18b \citep{fu22} and WASP-96b \citep{radica23}. As part of the Early Release Science program, \cite{feinstein23} presented the first NIRISS exoplanet transmission spectrum of WASP-39b. A single transit observation of the Saturn-mass exoplanet WASP-39 b showed clear water features in multiple bands, the potassium doublet, and evidence of clouds. NIRISS was also used to observe the rocky exoplanet Trappist-1 b \citep{lim23}, which displayed signatures of stellar contamination in its transmission spectra. A follow-up study of NIRISS/SOSS observations of WASP-96b presented in \cite{radica23} was able to constrain the abundances of \ch{H2O}, \ch{CO2}, and \ch{K} in its atmosphere \citep{taylor23}. \cite{madhusudhan23} combined NIRISS/SOSS and NIRSpec/G395H transmission spectra of the habitable zone sub-Neptune K2-18 b, and found strong detections of \ch{CH4} and \ch{CO2}, but a lack of \ch{H2O}, contradicting the original interpretations from this planet's WFC3 spectrum. \cite{radica24} found muted spectral features in the atmosphere of the hot Neptune LTT 9779 b using NIRISS/SOSS, suggesting either an extremely high-metallicity atmosphere or high-altitude clouds. \cite{fournier-tondreau24} re-analysed the ERO NIRISS/SOSS transmission spectrum for the Saturn-mass planet HAT-P-18 b, and demonstrated that the wavelength coverage provided by this data enables one to break the degeneracy between a cloudy atmosphere and one with a high metallicity, as well as distinguish between water features due to the planet's atmosphere and those from unocculted star-spots. \cite{benneke24} used NIRISS/SOSS and NIRSpec/G395H observations of the two-earth-radius exoplanet TOI-270 d to detect signals of \ch{CH4}, \ch{CO2} and \ch{H2O}, as well as potential signs of \ch{SO2} and \ch{CS2}. The abundances of multiple molecules allowed them to constrain the mean-molecular weight of the atmosphere, and infer the metal mass fraction. \cite{cadieux24} presented two transits of LHS 1140 b, a planet in the radius valley, and found evidence of unocculted faculae, with possible signs of Rayleigh scattering from an \ch{N2}-dominated atmosphere on the planet. 

Over the past few years, NIRISS/SOSS has proven highly successful in characterising the atmospheres of transiting exoplanets, enabling precise measurements of water features, detections of other molecules and clouds, and showing clear signatures of stellar activity. Of all the instruments on JWST, NIRISS/SOSS is the only one designed specifically for obtaining spectra of exoplanets. Whilst an increase in constraining power of this state-of-the-art instrument over WFC3 is to be expected, the extent of this improvement is unclear. This motivates our comparative study of the capabilities of NIRISS/SOSS and WFC3 for exoplanet characterisation.

\subsection{A case study of WASP-39b}

As part of the JWST Transiting Exoplanet ERS program \citep{stevenson16,bean18}, the hot gas giant exoplanet WASP-39 b was chosen to be observed with four of JWST's instruments: NIRCam (F322W2 filter, 2.4--\SI{4.0}{\um}, $R\sim2000$) \citep{ahrer23}, NIRISS/SOSS (0.6--\SI{2.8}{\um}, $R\sim1000$) \citep{feinstein23}, NIRSpec G395H  (2.87--\SI{5.14}{\um}, $R\sim2700$) \citep{alderson23}, and NIRSpec PRISM (0.6--\SI{5.3}{\um}, $R\sim100$) \citep{rustamkulov23}. WASP-39 b proved an ideal target for the ERS program due to its low density, providing large features in the transmission spectrum, and previous \ch{H2O} detection \citep{wakeford18}, as well as optimal timing with the observations. The ERS data provided a first look for the community at the capabilities of each instrument, and immediately revealed new insights into the atmosphere of WASP-39 b, such as absorption from \ch{CO2} \citep{jwsttransitingexoplanetcommunityearlyreleasescienceteam23} and the first detection of \ch{SO2} \citep{tsai23}.

Previous analyses of the pre-JWST spectra of WASP-39 b had already revealed its large spectral features, such as the strong sodium and potassium absorption lines with pressure broadened wings \citep{fischer16,sing16,nikolov16}, which provided evidence of a predominantly clear atmosphere. Analyses of the infrared spectra of WASP-39 b offered a range of possible \ch{H2O} abundances and metallicities. \cite{barstow17} used STIS and \textit{Spitzer} observations, and obtained an upper limit on the water abundance of about $1\times$ solar. \cite{wakeford18} used a combination of data from WFC3, STIS, the Very Large Telescope's (VLT) Focal Reducer and low dispersion Spectrograph (FORS2) instrument, and \textit{Spitzer}, and constrained the water abundance to $100$--$200\times$ solar. Using only the WFC3 data, \cite{fisher18} found a water abundance of 4--$24\times$ solar. \cite{welbanks19} used the STIS, WFC3 and \textit{Spitzer} data to constrain the water abundance to 4--$13\times$ solar. \cite{tsiaras18} re-reduced the WFC3 data and found a substantially reduced water abundance of 0.0001--$0.01\times$ solar. \cite{pinhas19} again used the combination of STIS, WFC3 and \textit{Spitzer} to constrain the water abundance to 0.01--$0.05\times$ solar. \cite{min20} also used these instruments and found a water abundance of 2--$50\times$ solar. \cite{kirk19} further combined these datasets with a spectrum from the William Herschel Telescope, and recovered a metallicity of 224--$347\times$ solar. This wide array of values highlights the possible uncertainty in constraining abundances from space-based data pre-JWST, and stresses the potential information to be gained from the increase in precision, resolution, and wavelength coverage of the JWST instruments. The current study provides an in-depth analysis of the information content of JWST's NIRISS instrument in comparison with WFC3, and determines the driving factor behind the improved constraints on a planet's atmospheric properties in the JWST era.

\subsection{Layout of the study}

In Section \ref{sec:methods} we discuss the methods we use, including the origin of the datasets and the atmospheric retrieval codes and their respective setups. In Section \ref{sec:results} we analyse our results across the various retrievals, comparing the constraints from WFC3 and NIRISS, and the effect of adding STIS. Additionally we compare the results from the WFC3 retrievals with retrievals on NIRISS spectra cut to the same wavelength range. We also perform mock retrievals on simulated data, in order to test our conclusions. In Section \ref{sec:discussion} we discuss the implications of our results, and make recommendations for future best practices. Finally, in Section \ref{sec:conclusion} we summarise our conclusions.

\section{Methods}
\label{sec:methods}

\subsection{Data}
\label{sec:data}

In this study we consider the transmission spectra of the hot-Jupiter WASP-39 b obtained from HST/WFC3, HST/STIS, and JWST NIRISS/SOSS. The WFC3 and STIS data are from \cite{wakeford18} and \cite{sing16,fischer16}, respectively, whilst the NIRISS data comes from the Exoplanet Early Release Science Team \citep{feinstein23}. For consistency with \cite{carter24} we use the \texttt{supreme-SPOON}\footnote{This pipeline is now known as \texttt{exoTEDRF} \citep{radica24a}} reduction from \cite{feinstein23}. As described in \citet{feinstein23}, the raw, uncalibrated data frames were reduced using a combination of steps from the default \texttt{jwst} calibration pipeline and custom routines, including a group-level correction of time-variable 1/$f$ noise as is commonly performed for JWST's other instruments \citep[e.g.,][]{alderson23, rustamkulov23}. The stellar spectra were extracted from the 2D data frames using the \texttt{ATOCA} and \texttt{APPLESOSS} algorithms \citep{darveau-bernier22, radica22} to explicitly model the overlap of the first and second diffraction orders on the detector. The light curves were fit at the pixel-level (that is, one light curve per detector column) using the \texttt{juliet} library \citep{espinoza19}. Along with the \texttt{BATMAN} transit model, we included in the fits a systematics model consisting of a linear trend with the $x$-position of the spectral trace on the detector. The fitted transit depths were then binned to a constant $R=100$.

These datasets were reduced using different techniques, pipelines, and prior assumptions, which could lead to inhomogeneities such as offsets across the instruments \citep{mugnai24}. However, we remain agnostic to these differences, in an effort to highlight some of the issues involved with combining spectra from different papers, when a homogeneous reanalysis of the raw data may not be possible. Furthermore, even in the case of homogeneously analysed data, differences can still arise, such offsets between instruments \citep{madhusudhan23,welbanks24,benneke24}, changes in stellar activity between epochs \citep{rosich20}, or variability in the planetary atmosphere \citep{morello23,parviainen23,changeat24}.

\begin{figure*}
    \centering
    \includegraphics[width=\textwidth]{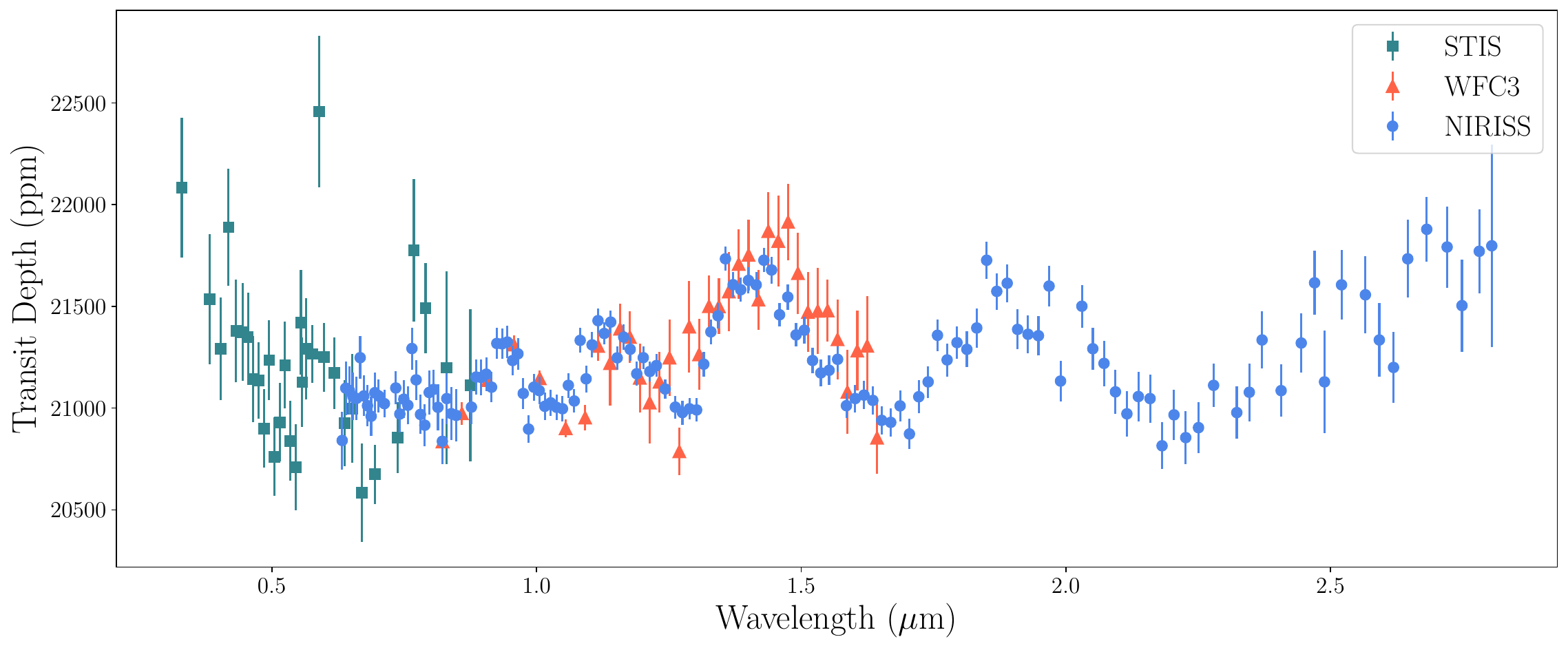}
    \caption{STIS, WFC3 and NIRISS transmission spectra for WASP-39 b. The data are plotted as they are published in \protect\cite{sing16,fischer16} (STIS), \protect\cite{wakeford18} (WFC3), and \protect\cite{feinstein23} (NIRISS), with no offsets added.}
    \label{fig:wasp39b_data}
\end{figure*}

Figure \ref{fig:wasp39b_data} shows the three transmission spectra for WASP-39 b. Whilst spectral features are clearly visible in all three spectra, there are some obvious discrepancies between the datasets. Firstly, the WFC3 data has a slight offset from the NIRISS data at the \SI{1.4}{\um} water feature. Secondly, many of the STIS data points appear inconsistent with the blue end of the NIRISS data, although the error bars on these are large. 

\subsection{Retrieval setup}
\label{sec:retrieval_setup}

In order to perform a comparison across all three datasets, we apply atmospheric retrievals to these spectra. This involves using a Bayesian sampling algorithm, such as nested sampling or an MCMC, coupled with an atmospheric model, to search parameter space for the values that could explain the spectroscopic observations \citep[e.g.,][]{madhusudhan09,benneke13}. There are many choices involved in building a retrieval code, such as the model assumptions, molecular line-lists, and sampling method. We test two different retrieval codes in this work: \texttt{BeAR} (formerly known as \texttt{Helios-r2}) \citep{kitzmann20}, and \texttt{CHIMERA} \citep{line13}. For a comprehensive table of current atmospheric retrieval codes for exoplanets see \cite{macdonald23}.

\subsubsection{\texttt{BeAR}}
\label{sec:heliosr2}

\texttt{BeAR}\footnote{The open-source code \texttt{Helios-r2} has recently been superseded by a new version that has been renamed to \texttt{BeAR}, the Bern Atmospheric Retrieval code, which is used in this study. This code can be found here: \url{https://github.com/newstrangeworlds/bear}.} is a one-dimensional, open-source, GPU-accelerated atmospheric retrieval code \citep{kitzmann20}, using nested sampling via the \texttt{MultiNest} package \citep{feroz08,feroz09}. It was originally developed for analysing the emission spectra of brown dwarfs, to which it has been applied extensively \citep{lueber22}. It has since been updated for exoplanets, for both transmission and secondary eclipse spectra. \texttt{BeAR} uses line-by-line opacity sampling to describe the wavelength-dependent absorption and scattering. In this work, we include the following molecules and their associated \texttt{ExoMol} and \texttt{HITEMP} line-lists: \ch{H2O} \citep{polyansky18}, \ch{CO} \citep{li15,somogyi21}, \ch{CO2} \citep{rothman10}, \ch{CH4} \citep{yurchenko13,yurchenko14}. Additionally, we include the alkalis \ch{Na} and \ch{K}, for which we use the Kurucz line lists \citep{kurucz95}. The pressure broadening of their strong resonance lines are calculated with data from \cite{allard19} and \cite{allard16}, see \citet{kitzmann20} for details. The opacities are calculated using \texttt{Helios-k} \citep{grimm15,grimm21}, at a resolution of \SI{0.01}{\per\cm}, and can be found on the DACE opacity database\footnote{\url{https://dace.unige.ch}}. In this study, the models are sampled at a resolution of \SI{1.0}{\per\cm}. We assume the atmosphere is hydrogen-dominated, with a solar \ch{He}/\ch{H2} ratio of 0.17 \citep{asplund09}. We include collision-induced absorption from \ch{H2}-\ch{H2} \citep{abel11} and \ch{H2}-\ch{He} \citep{abel12}, using the \texttt{HITRAN} database, as well as Rayleigh scattering due to \ch{H2} \citep{vardya62}. A plot of all the opacity sources used by \texttt{BeAR} in this study is shown in Figure \ref{fig:opacities}. 

\texttt{BeAR} can be run using the chemical equilibrium model \texttt{FastChem} \footnote{The open-source code \texttt{FastChem} can be found here: \url{https://github.com/newstrangeworlds/FastChem}.} \citep{stock18,stock22,kitzmann23}, or using free chemistry, with independently retrieved molecular volume mixing ratios $X$. We opt for the latter, in order to avoid assumptions about the chemistry of the atmosphere. \texttt{BeAR} also has the option to include variable abundance profiles for the chemical species, though we do not use this. The atmosphere in \texttt{BeAR} is divided into equal layers in log-pressure space. We use a top pressure of $10^{-6}$ bar and a bottom pressure of $10$ bar, with 200 layers in between. This bottom, or ``reference", pressure corresponds to the level of the planet radius $R_p$.

When considering clouds, \texttt{BeAR} has the option to include either a grey or non-grey cloud layer. For the grey-cloud, one retrieves a cloud-top pressure $P_{\rm cloudtop}$, a cloud optical depth $\tau_{\rm cloud}$, and a cloud-bottom pressure $P_{\rm cloudbottom}$, set by a parameter $b_c$ such that $P_{\rm cloudbottom}=b_cP_{\rm cloudtop}$. $b_c$ effectively sets the size of the cloud layer. In the non-grey cloud model, one still retrieves the cloud top and bottom, but instead of a constant optical depth, we now have an equation depending on wavelength,
\begin{equation}
    \tau(\lambda) = \tau_{\rm ref} \frac{Q_0x_{\lambda_{\rm ref}}^{-a_0} + x_{\lambda_{\rm ref}}^{0.2}}{Q_0x_{\lambda}^{-a_0} + x_{\lambda}^{0.2}},
\end{equation}
where $x_\lambda=2\pi r_{\rm cloud}/\lambda$. The additional free parameters are then $Q_0$ (a proxy for the cloud composition), $a_0$ (the power law index in the small particle limit), and $r_{\rm cloud}$ (the cloud particle radius). For a detailed explanation of this cloud model, see \citet{kitzmann18} or \citet{lueber22}.

In addition to the atmospheric parameters mentioned above, \texttt{BeAR} also retrieves the planet and star parameters, such as the surface gravity $\log{g}$, the planet radius $R_p$ at the bottom of the assumed atmospheres, and the radius of the host star $R_*$. For the purposes of this study, we fix the surface gravity and stellar radius ($\log{g}=2.63$, $R_*=0.9R_\odot$ \citep{faedi11}), and only retrieve the planet radius at a pressure of 10 bar. The priors used for all the possible model parameters are found in Table \ref{tab:priors}. 

\begin{table}
    \centering
    \begin{tabular}{cccc}
         Parameter & Unit & Prior Distribution & Limits \\
         \hline
         $R_p$ & $R_{\rm J}$ & Uniform & $[1.1, 1.4]$ \\
         $X_{\ch{H2O}}$ & -- & Log-uniform & $[10^{-12}, 1.0]$ \\
         $X_{\ch{CO}}$ & -- & Log-uniform & $[10^{-12}, 1.0]$ \\
         $X_{\ch{CO2}}$ & -- & Log-uniform & $[10^{-12}, 1.0]$ \\
         $X_{\ch{CH4}}$ & -- & Log-uniform & $[10^{-12}, 1.0]$ \\
         $X_{\ch{Na}}$ & -- & Log-uniform & $[10^{-12}, 1.0]$ \\
         $X_{\ch{K}}$ & -- & Log-uniform & $[10^{-12}, 1.0]$ \\
         $T$ (isothermal) & K & Uniform & $[500, 3000]$ \\
         $T_{\rm irr}$ (non-isothermal) & K & Uniform & $[500, 3000]$ \\
         $\kappa_{\rm IR}$ & cm$^2$g$^{-1}$ & Log-uniform & $[10^{-4}, 10]$\\
         $\gamma$ & -- & Log-uniform & $[10^{-2}, 100]$\\
         $\tau_{\rm cloud}$ (grey) & -- & Log-uniform & $[10^{-5}, 20]$ \\
         $\tau_{\rm ref}$ (non-grey) & -- & Log-uniform & $[10^{-5}, 20]$ \\
         $Q_0$ & -- & Uniform & $[1, 100]$ \\
         $a_0$ & -- & Uniform & $[3, 6]$ \\
         $r_{\rm cloud}$ & cm & Log-uniform & $[10^{-7}, 10^{-1}]$ \\
         $P_{\rm cloudtop}$ & bar & Log-uniform & $[10^{-5}, 10]$ \\
         $b_c$ & -- & Log-uniform & $[1, 10^3]$ \\
         \hline 
    \end{tabular}
    \caption{Prior distributions for all possible model parameters for \texttt{BeAR}.}
    \label{tab:priors}
\end{table}

\subsubsection{\texttt{CHIMERA}}
\label{sec:chimera}

We use \texttt{CHIMERA}\footnote{The open-source \texttt{CHIMERA} code can be found here: \url{https://github.com/mrline/CHIMERA}} to perform both free and chemically consistent spectral retrievals. \texttt{CHIMERA} is the only framework in this study that uses the correlated-$k$ approach \citep{lacis91} when computing transmission through the atmosphere. The $k$-tables are computed at a resolution of R=1000; the line-by-line data used to calculate the $k$-tables are from the following sources: H$_2$O \citep{polyansky18, freedman14}, CO$_2$ \citep[][]{freedman14}, CO \citep{rothman10}, CH$_4$ \citep{rothman10},  Na \citep[]{kramida18, allard19}, and K \citep[]{kramida18, allard16}, and were computed following the methods described in \citet{gharib-nezhad21, grimm21}. Again, we assume the atmosphere is dominated by H$_2$, with a solar He/H$_2$ ratio of 0.1764; therefore, we also model the H$_2$-H$_2$ and H$_2$-He collision-induced absorption (CIA) \citep[][]{richard12}. The \texttt{CHIMERA} framework has been extensively used to analyse the transmission spectra of exoplanets obtained with HST/WFC3 and JWST \citep[e.g.][]{kreidberg14a,taylor23} and is part of the cutting edge frameworks being used in the Early Release Science collaboration \citep{feinstein23}.

In \texttt{CHIMERA}, we treat the inclusion of clouds in three ways: no cloud, grey cloud or non-grey cloud. For our grey cloud we fit for a cloud top pressure, at pressures larger than this we set the atmosphere to be completely opaque. For our non-grey model we employ the \citet{ackerman01} cloud model to produce realistic vertical droplet profiles over a broad range of droplet sizes (for a specific condensate), given a sedimentation efficiency (f$_{\text{sed}}$), eddy diffusivity (K$_{zz}$ cm$^2$/s), cloud base pressure, and the condensate mixing ratio at the cloud base. Here we assume enstatite (MgSiO$_3$) physical/optical properties for the condensate as a representative non-grey cloud at short wavelengths, following the conclusions of \cite{gao20}. However, the cloud top pressure is a separate free parameter, not linked with the condensation curve of enstatite. 

\subsubsection{Temperature-Pressure Profiles}

Previous studies have shown that, even in the case of transmission spectra, non-isothermal profiles can be required to accurately retrieve the molecular abundances from JWST data \citep[e.g.,][]{rocchetto16}. This motivates us to test both isothermal and non-isothermal retrievals in our study. Both \texttt{BeAR} and \texttt{CHIMERA} have the option to include different temperature-pressure parameterisations, and for this study we adopt the "Guillot" profile \citep{guillot10,parmentier14}. In this work, the parameterisation follows \cite{guillot10} (the semi-grey case), where the parameters are the infrared opacity $\kappa_{\rm IR}$, the irradiation temperature $T_{\rm irr}$, the internal temperature $T_{\rm int}$, the ratio between optical and IR opacity $\gamma$, and the redistribution factor $f$. In our retrievals, we fix the internal temperature to $200$K, and $f$ to 0.25 to correspond to an averaging of the incoming flux over the planetary surface. This leaves three free parameters, the priors for which are shown in Table \ref{tab:priors}. 

\subsubsection{Model Complexity}

For each of our retrieval codes, we test a range of different retrieval models, with increasing complexity. We start with an isothermal, cloud-free model, then add a grey-cloud and a non-grey cloud or haze, then do the same for the non-isothermal model. This leads to six different models to test for each individual dataset and various combinations, for each code. It it worth noting that the two codes have different parameterisations for the non-grey cloud model. Although the ERS study of the NIRISS spectra of WASP-39b found evidence for spacially inhomogeneous cloud cover in the planet's atmosphere \citep{feinstein23}, we opt not to include this level of complexity in our model, in order to limit the number of test cases, and because the wavelength region driving this model selection (2.0--2.3 microns) is not encompassed in the WFC3 range. 

The Bayesian Evidence, obtained from a nested sampling retrieval, can be compared to determine when one model is more favoured by the data than another. The Bayesian Evidence penalises the addition of parameters that do not sufficiently improve the fit to the data, thus providing a method of formally implementing Occam's Razor. \cite{trotta08} gives the difference in log Bayesian Evidence (or the logarithm of the Bayes factor, $\ln B_{ij}$) corresponding to inconclusive, weak, moderate, and strong evidence for model $i$ over model $j$ as $\ln B_{ij}<1.0$, $1.0 < \ln B_{ij} < 2.5$, $2.5 < \ln B_{ij} < 5.0$, and $\ln B_{ij} > 5.0$, respectively.

\section{Results}
\label{sec:results}

\subsection{\texttt{BeAR} vs \texttt{CHIMERA}}
\label{sec:heliosr2_vs_chimera}

\begin{figure*}
    \centering
    \includegraphics[width=\textwidth]{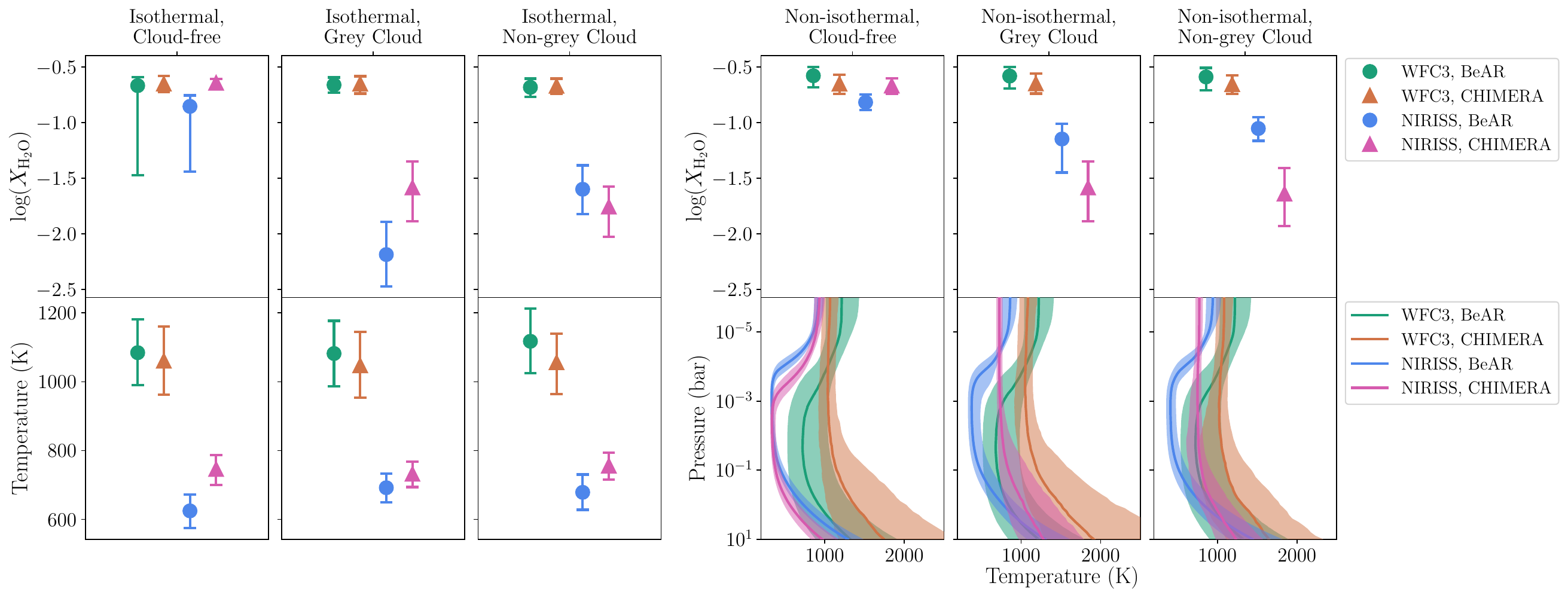}
    \caption{A summary of the retrieval results for the different retrievals applied to the WFC3 and NIRISS data using \texttt{BeAR} and \texttt{CHIMERA}. The symbols and errorbars denote the median and 1-$\sigma$ values from the posterior distributions. The top row shows the retrieved water abundances for each case. The bottom row shows the retrieved temperatures, where the left three panels display single isothermal temperatures and the right three panels show the full T-P profiles for the non-isothermal cases. It can be seen that the two codes agree to within 1-sigma for the majority of water abundances and isothermal temperatures. The non-isothermal temperatures slightly deviate owing to the differences in the codes, such as cloud parameterisation}
    \label{fig:code_comparison}
\end{figure*}

In this section we compare results from our two retrieval codes. Figure \ref{fig:code_comparison} shows a summary of the retrieved water abundances and temperatures for each of the codes applied to the WFC3 and NIRISS data for the different models. The codes generally agree to within 1- or 2-$\sigma$, with the largest disagreement occurring for the most complex model (non-isothermal + non-grey cloud), applied to the NIRISS data. Given the similarity of the T-P profiles in the cloud-free case for NIRISS, this disagreement is likely due to the different parameterisations of clouds/hazes (see Sections \ref{sec:heliosr2} and \ref{sec:chimera}). Though the two codes use different line lists for some of the species, this is unlikely to cause major differences across the WFC3 and NIRISS wavelengths, since the water line list is consistent between the two, and this is the main absorber in this spectral region. These discrepancies in the codes' results highlight the sensitivity of retrievals to minor differences in the models, particularly for data with the precision of JWST. 

Our study focuses on a comparison of the different types of data, rather than aiming to make confident claims about the atmosphere of WASP-39 b. Therefore, in the rest of the paper we will only use \texttt{BeAR} to perform the retrievals to save additional computations. However, this code comparison demonstrates the need to compare results from multiple retrieval codes with different assumptions to ensure any claims are not model dependent. This is particularly important when analysing JWST data, due to its increased precision and sensitivity. For a comprehensive comparison of different retrieval codes applied to JWST data, see Welbanks et al., \textit{in prep}.

\subsection{WFC3 vs NIRISS}
\label{sec:wfc3_vs_niriss}

\begin{figure*}
    \centering
    \includegraphics[width=0.47\textwidth]{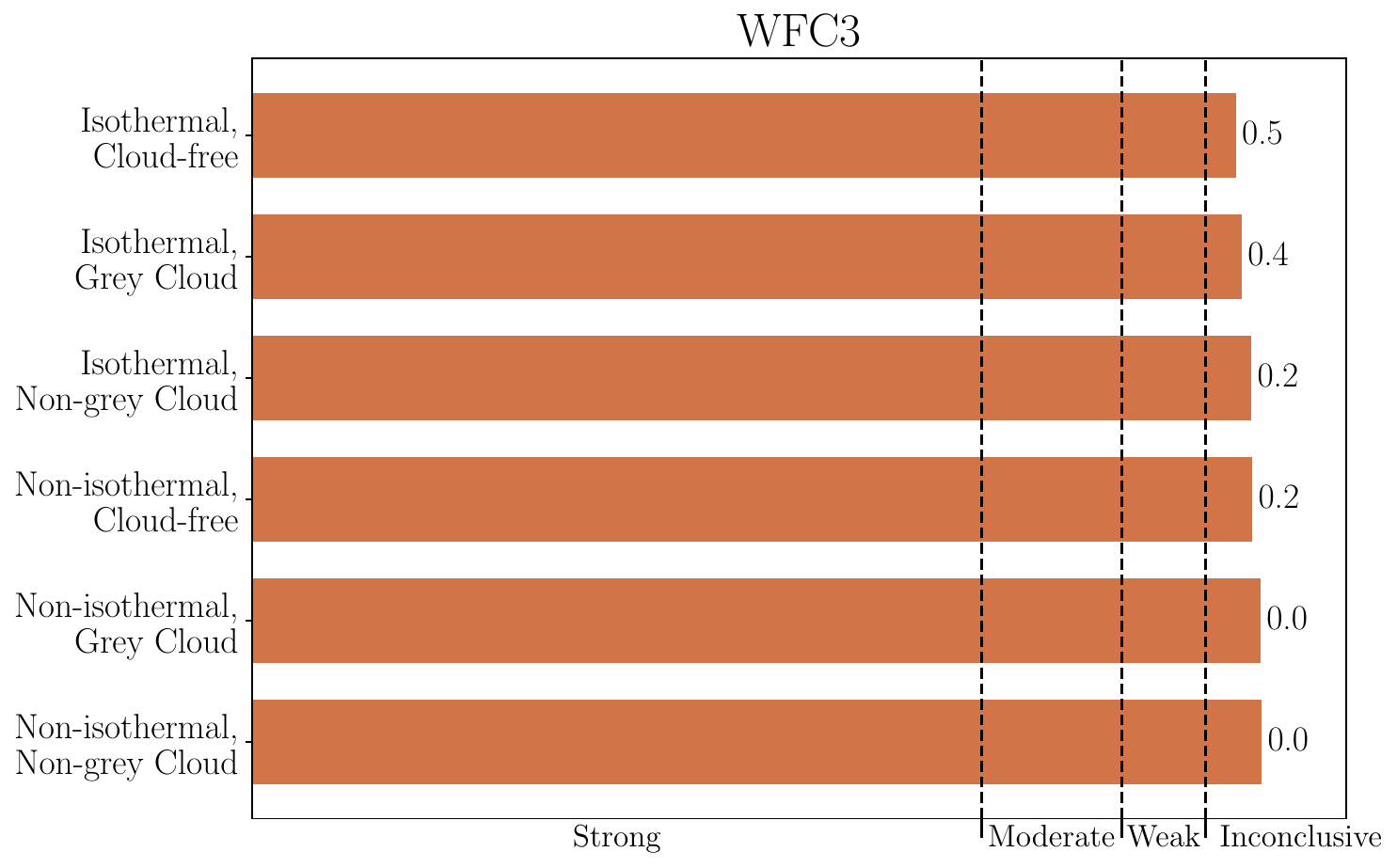}
    \hspace{5mm}
    \includegraphics[width=0.47\textwidth]{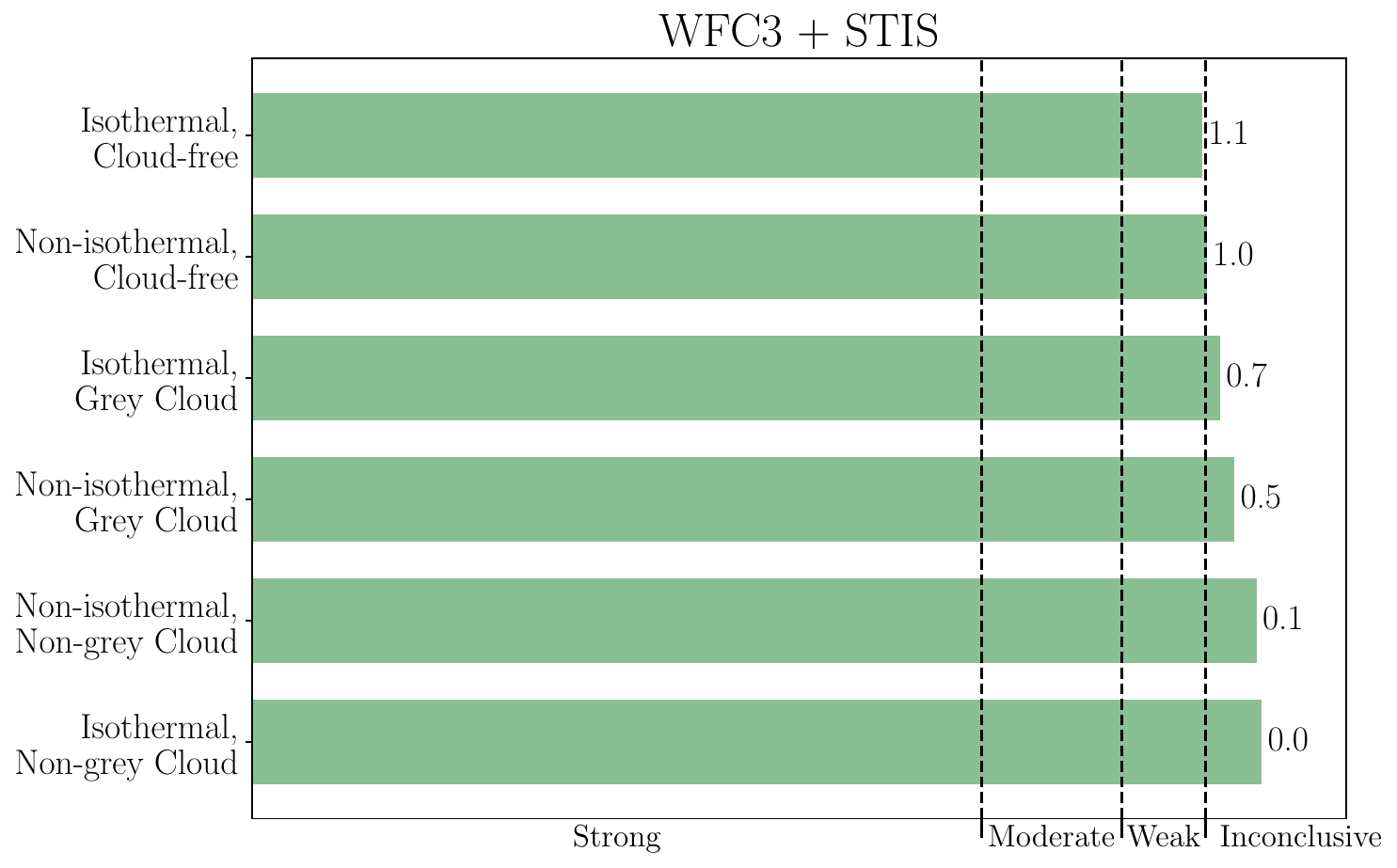}\\
    \vspace{5mm}
    \includegraphics[width=0.47\textwidth]{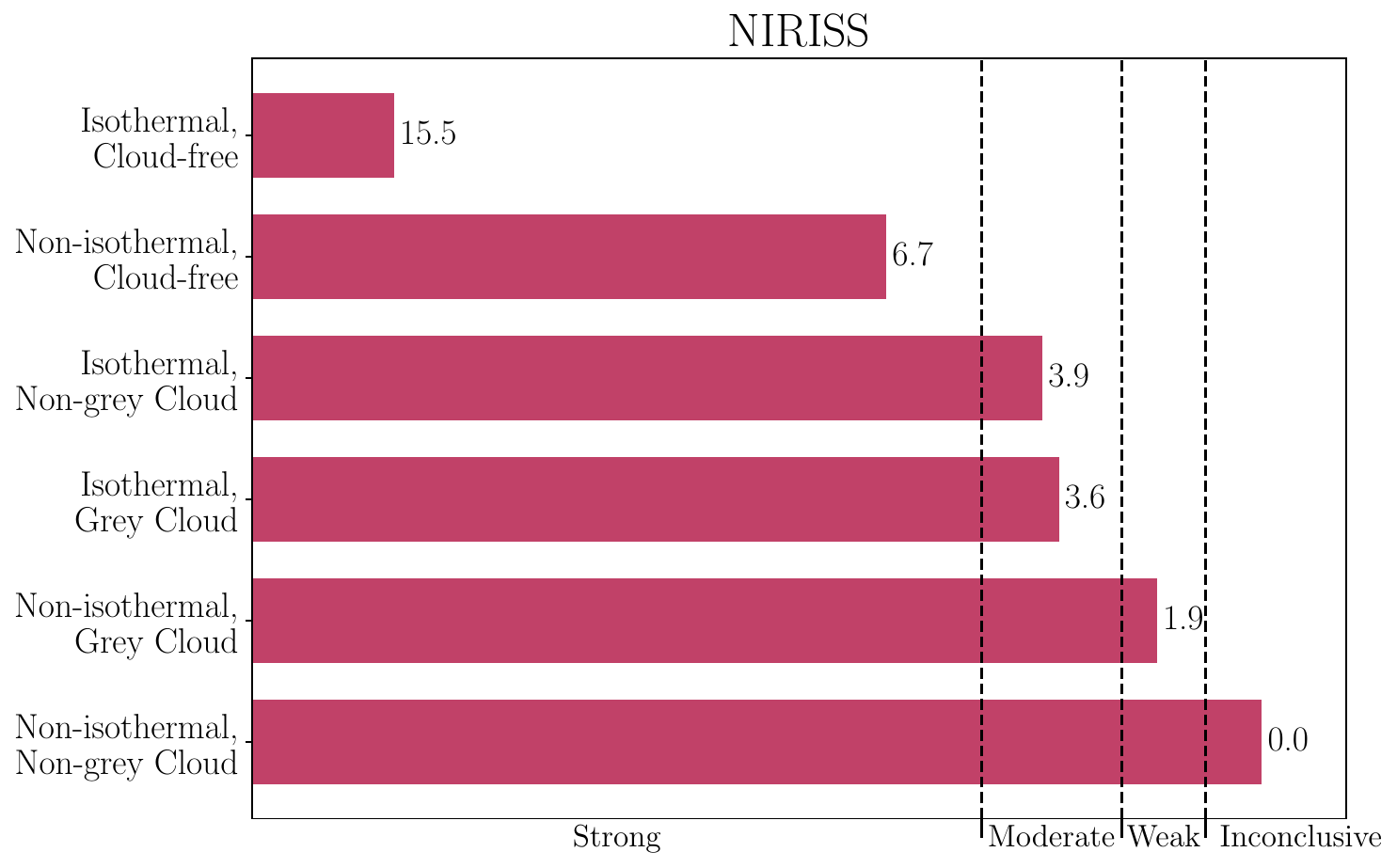}
    \hspace{5mm}
    \includegraphics[width=0.47\textwidth]{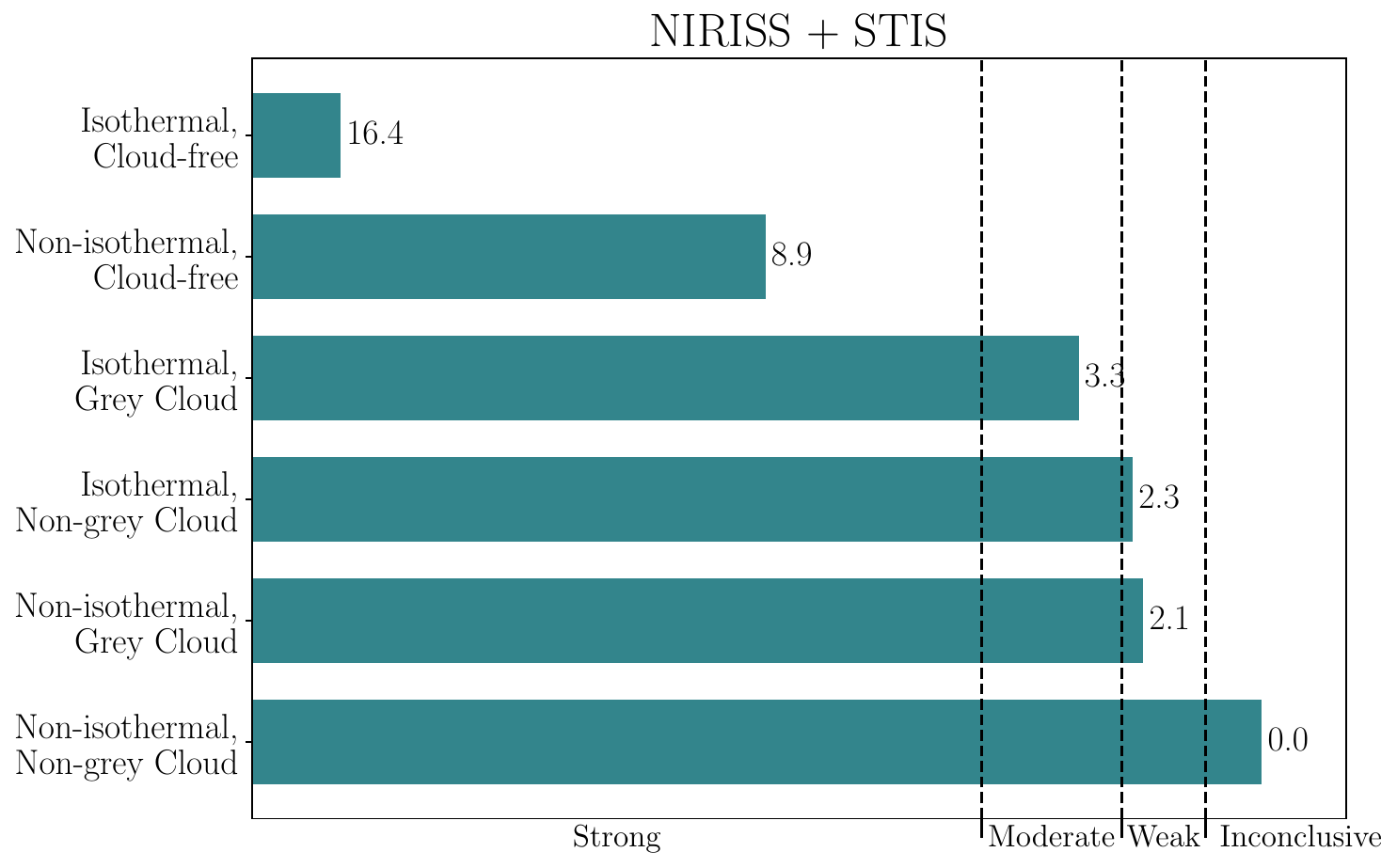}
    \caption{Bayesian evidence hierarchy for the different models used in each retrieval, for each data set. The models are sorted by Bayesian evidence, with the most favoured model at the bottom. Bayes factor values are displayed next to the bars, comparing each model to the most favoured model. The $x$-axis shows the division between inconclusive, weak, moderate, and strong evidence for ruling out models with respect to the most favoured model. These boundaries correspond to significances of $2.1\sigma$ (weak), $2.7\sigma$ (moderate), and $3.6\sigma$ (strong).}
    \label{fig:Bayesian_Evidences}
\end{figure*}

\begin{figure*}
    \centering
    \includegraphics[width=\textwidth]{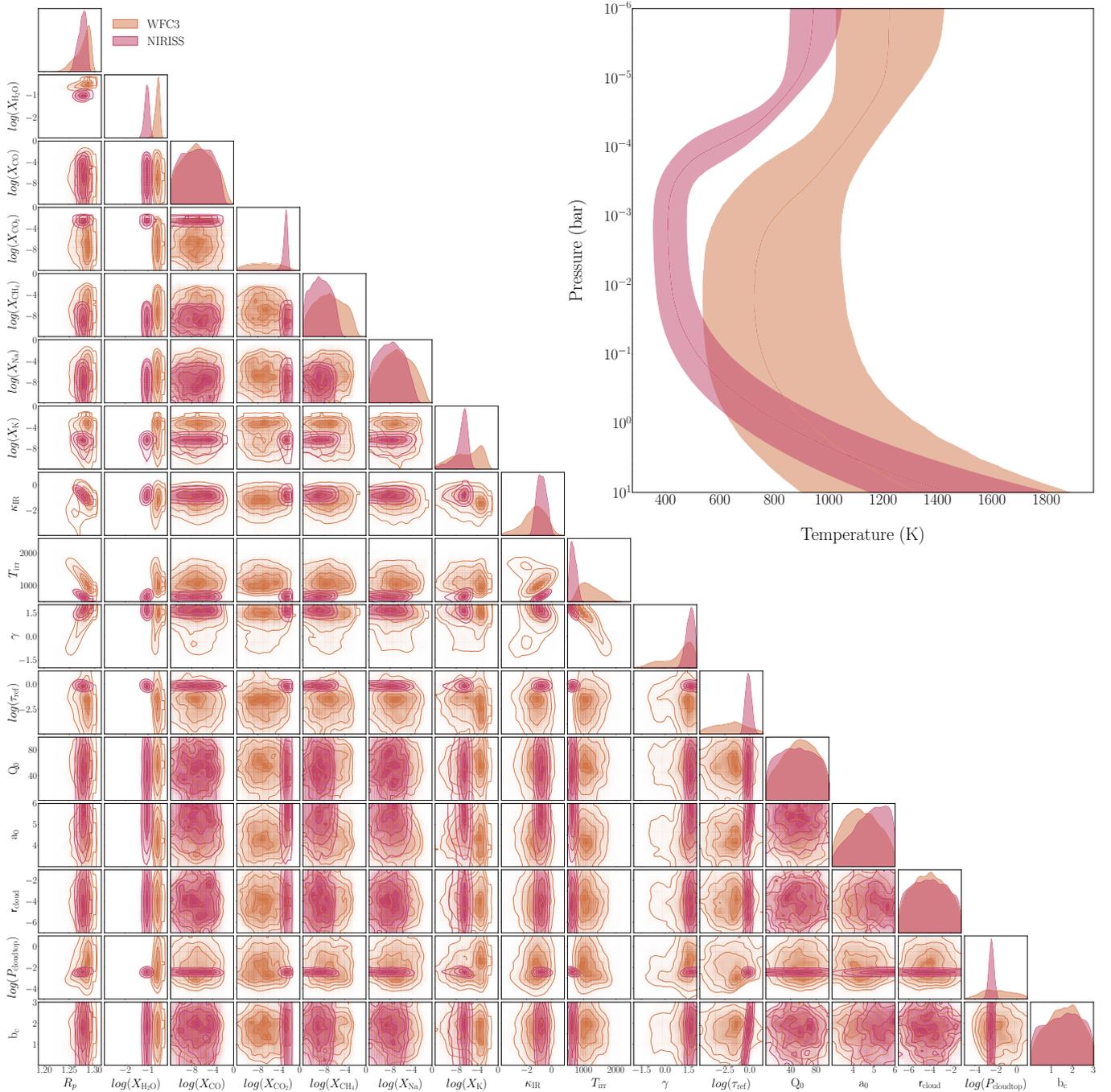}
    \caption{Retrievals using the non-isothermal + non-grey cloud model for the WFC3 (orange) and the NIRISS (red) data. The median and 1-sigma limits of the retrieved temperature-pressure profiles are shown in the top right corner.}
    \label{fig:WFC3_NIRISS_retrieval}
\end{figure*}

In this section we compare the results from the \texttt{BeAR} retrievals on the WFC3 and NIRISS data. The left-hand panels of Figure \ref{fig:Bayesian_Evidences} show the Bayesian evidence hierarchy for the retrievals on these individual datasets. In the WFC3 case, the Bayesian model comparison is inconclusive across all models, implying there is no real preference to fit this dataset. For NIRISS, on the other hand, the non-isothermal + non-grey cloud model is the best-fit model, whilst the other models are either weakly, moderately, or, in the case of the cloud-free models, strongly ruled out by the Bayesian evidence. With this in mind, Figure \ref{fig:WFC3_NIRISS_retrieval} compares the retrievals for the non-isothermal + non-grey cloud model for the WFC3 and NIRISS spectra.

There is a clear discrepancy between the water abundance posteriors from the two retrievals. WFC3 gives an extremely high value of $\log{X_{\ch{H2O}}}=-0.59^{+0.08}_{-0.12}$ corresponding to an atmosphere of $\sim25$\% water, whilst NIRISS gives a lower value of $\log{X_{\ch{H2O}}}=-1.05^{+0.10}_{-0.11}$. This is likely due a difference in shape of the water feature, seen in Figure \ref{fig:wasp39b_data}. Without the wider wavelength coverage, the WFC3 data holds little information about the continuum, providing a challenge for the retrieval to fit spectral features accurately. This high water abundance causes the mean molecular weight of the atmosphere to increase (to $\sim6.4$), causing a decrease in the atmospheric scale height, squashing spectral features and thus allowing an extremely high abundance to fit a moderately-sized water feature. Previous studies have shown that, in the first-order analytical case and when only one absorber is present, it is not possible to measure its abundance from the transmission spectra due to degeneracies \citep{lecavelierdesetangs08,benneke12,line16}, except in the case of very high abundance, due to the mean-molecular weight effect, or very low abundance, where the continuum due to CIA sets a minimum \citep{welbanks19}. This degeneracy is broken when the effects of temperature and pressure on the species' opacity is taken into account. However, these effects are quite weak, and the large error bars and low-resolution of WFC3 data make them hard to distinguish. This is consistent with our extremely high value for the water abundance retrieved from WFC3 data alone. The difference in shape of the water feature between WFC3 and NIRISS could be due to different systematics in the instruments, but the lack of additional features in WFC3 likely make the retrieval more sensitive to these changes. 

Since the narrow wavelength range of WFC3 restricts any detections of molecules to \ch{H2O} only, the other molecules unsurprisingly have unconstrained posteriors. In contrast, the NIRISS retrieval shows some evidence for \ch{CO2} and \ch{K}, and upper limits for \ch{CO}, \ch{CH4}, and \ch{Na}. These additional absorbers allow for abundance constraints not only at extreme values, as suggested by \cite{line16}. To compute the detection significances of the constrained species, we ran additional retrievals for the non-isothermal + non-grey cloud case, omitting each molecule in turn. For the WFC3 retrievals, we find a detection significance of 9.0$\sigma$ for \ch{H2O}. In comparison, the NIRISS retrievals obtain significances of 21.3$\sigma$ for \ch{H2O}, 2.8$\sigma$ for \ch{CO2}, and 2.1$\sigma$ for \ch{K}. This demonstrates a quantifiable gain in information in the NIRISS data when compared to WFC3, providing a more robust detection of water, as well as the ability to probe additional species.

For the clouds, the WFC3 retrieval shows unconstrained posteriors for every cloud parameters. This is unsurprising as the cloud-free models are equally as preferred by the Bayesian evidence, suggesting clouds are not detected. The posteriors spanning the entire prior range for each cloud parameter also indicates that clouds would not be detected in this data, even if they were present in the atmosphere. This is likely due to the extremely high retrieved water abundance, preventing any visible effect of a cloud. In contrast, the NIRISS retrieval shows a clear cloud detection, with a significant reference optical depth, and a cloud top pressure of $\sim$\SI{10}{mbar}, placing it in the region predicted to be probed by transmission spectroscopy. 

For the temperature-pressure profile, the WFC3 retrieval shows a fairly unconstrained profile, with some signs of an inversion in the upper atmosphere. In the NIRISS case, the profile appears to have a similar shape, but more tightly constrained and reaching cold temperatures in the middle of the atmosphere. These cold temperatures may not be realistic for a planet with an equilibrium temperature of 1100 K. The inversions seen at high altitudes are likely caused by the model requiring a hotter temperature to fit the alkalis (potassium in this case) high in the atmosphere. However, this model is not entirely consistent as it fits a pressure-dependent temperature alongside molecular abundances that are constant in altitude. It also assumes a single temperature profile across the entire terminator, neglecting variations in longitude and latitude. Therefore, these retrieved T-P profiles are unlikely to be an accurate representation of the vertical structure of the atmosphere. Furthermore, there is only moderate evidence favouring this model over several isothermal models, so this level of complexity is not necessarily required to fit the data. In general, WFC3 retrievals obtain higher temperatures than those of NIRISS, as shown in Figure \ref{fig:Retrieval_comparison}. The issues with the reliability of the temperature profiles represent a limitation of simplified 1D retrieval models that are commonly used in exoplanet analysis, which has been discussed in detail in, e.g., \cite{macdonald20}.

Our NIRISS retrievals give a water abundance of $\log{X_{\ch{H2O}}}=-2$ to $-1$, depending on the model used. In comparison to literature values, this is generally in good agreement with other studies of the JWST data. \cite{feinstein23} constrain the C/O ratio and metallicity of the planet by grid-fitting self-consistent equilibrium models, and find values that correspond to a log water abundance of $-1.86$. \cite{rustamkulov23} use similar methods to analyse the planet's atmosphere from the NIRSpec PRISM data, and find values corresponding to slightly lower log water abundances ($-3$ to $-2.5$). This deviation between results could be due to the assumption of chemical equilibrium and strong absorption features in the PRISM data from other molecules, such as \ch{CO2}. \cite{ahrer23} analyse the NIRCam data and find C/O ratios and metallicities that correspond to $\log{X_{\ch{H2O}}}=-3.3$ to $-1.4$. \cite{lueber24} retrieve a log water abundance of $-1.45$ to $-1.13$ from the NIRISS data, depending on the model complexity they use. Their analysis of the PRISM and NIRCam spectra however, also give lower abundances ($\log{X_{\ch{H2O}}}=-3.53$ to $-3.10$ and $\log{X_{\ch{H2O}}}=-4.15$ to $-4.10$, respectively). Even more depleted \ch{H2O} abundance values are found from their analysis of the NIRSpec G395H data, but they discuss how the lack of full coverage of the water feature combined with a flexible non-grey cloud could be biasing those results. Another study of the PRISM spectrum of WASP-39b, \cite{constantinou23}, also found a lower water abundance ($\log{X_{\ch{H2O}}}=-4.85$ to $-3.14$). However, they only use the 3--5 micron range of the PRISM data, and combine it with WFC3. Although not JWST data, \cite{wakeford18} use a combination of WFC3, STIS, VLT and \textit{Spitzer} data, and find a log water abundance of $-1.37$ from their free chemistry retrievals. On the whole, this demonstrates the wide range of \ch{H2O} abundances that are still consistent with the data, even in the era of JWST.

\subsection{Adding STIS}
\label{sec:adding_stis}

In this section we look at the effect of adding the STIS data to each of the NIRISS and WFC3 spectra in turn. Again, we stress that the different reduction techniques could cause discrepancies between the data, but since we wish to test potential effects of this we do not re-reduce the data before combining them.

The Bayesian evidence for these retrievals is shown in the two right-hand panels of Figure \ref{fig:Bayesian_Evidences}. For WFC3 + STIS, we are now able to weakly rule out the cloud-free models, as the STIS data has widened the wavelength coverage and given us some information about the bluer points, where clouds have their strongest effect. However, the rest of the models are all comparable in terms of the evidence. For NIRISS + STIS, the evidence is similar to the NIRISS-only case, favouring the non-isothermal + non-grey cloud case, and strongly ruling out cloud-free models. 

\subsubsection{Water and Temperature}
\label{sec:addingstis_watertemp}

\begin{figure*}
    \centering
    \includegraphics[width=\textwidth]{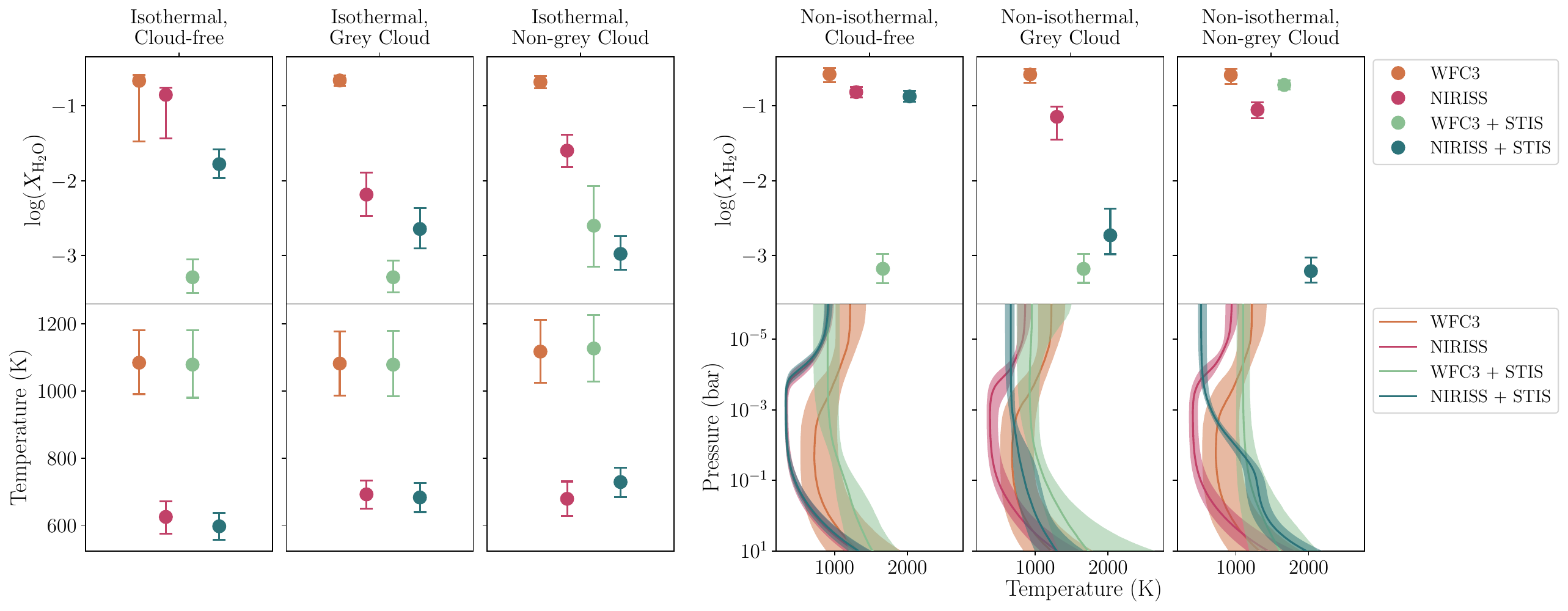}
    \caption{Summary of the retrieval results for the water abundance and temperature, for all the models and all the data set combinations.}
    \label{fig:Retrieval_comparison}
\end{figure*}

To compare the effect of adding STIS across all retrievals, Figure \ref{fig:Retrieval_comparison} shows a summary of the retrieved water abundance and temperature for every model. For \ch{H2O}, adding STIS to WFC3 almost always brings the abundance down substantially, because it adds more information about the continuum and because the large features from \ch{Na} and \ch{K}, along with the visible optical slope (see Figure \ref{fig:wasp39b_data}), prevent a high mean-molecular weight atmosphere from fitting the spectrum. For the non-isothermal + non-grey cloud model, the cloud top pressure is driven to unphysically high values (see Figure \ref{fig:WFC3_NIRISS_STIS_retrieval}), hitting the edge of the prior at $10^{-5}$ bar (one order of magnitude below the top of the atmosphere in the model). In contrast, the retrieval obtains a low value for the cloud optical depth ($\tau_{\rm ref}$), balancing the high cloud top pressure. This is likely due to the extremely steep slope on the edge of the STIS data. 

For NIRISS, adding STIS also brings the water abundance down, though with varying impact. For the isothermal models, adding STIS to either WFC3 or NIRISS has little-to-no effect on the retrieved temperature. For the non-isothermal models, adding STIS to WFC3 seems to increase the temperature at the bottom of the atmosphere and decrease it at the top, removing the temperature inversion, and favouring a more isothermal profile. For NIRISS, the same effect appears for the cloudy models, but no change occurs when adding STIS for the cloud-free case. The smaller effect of adding STIS in the cloud-free cases can be explained by the model's inability to adjust the fit to the STIS data without the cloud. When considering the non-isothermal cases, it is worth noting that we do not expect temperature profiles to be robustly retrieved from transmission spectra due to their narrow probed-pressure range, and the fact that the temperature only enters into the equation in two places -- the scale height and the temperature-dependent opacities. The latter is a weak effect at low resolution, so the temperature profile constraints rely on probing different scale heights across the spectrum. Despite this, some of the T-P profiles appear very tightly constrained, even for altitudes at which there is no information from the transmission spectrum. It is possible these models are over-fitting the data. It is also likely that the choice of T-P profile parametrisation will have a large effect on the results \citep{blecic17}. Furthermore, since our retrievals assume constant mixing ratios with altitude, the temperature could be compensating for varying mean molecular weight in the atmosphere, though we do not explore this. 

As with the WFC3 and NIRISS retrievals, we also run additional retrievals on the datasets combined with STIS, omitting particular species in order to compute their detection significances. For WFC3 + STIS, we find significances of 8.8$\sigma$ for \ch{H2O}, 3.8$\sigma$ for \ch{Na}, and 2.6$\sigma$ for \ch{K}. In comparison, for NIRISS + STIS, we obtain significances of 21.3$\sigma$ for \ch{H2O} and 2.3$\sigma$ for \ch{Na}. Interestingly, when \ch{K} is removed the Bayesian evidence of the retrieval improves, despite it having a well-constrained posterior (see Figure \ref{fig:WFC3_NIRISS_STIS_retrieval}). Possibly the overlapping \ch{K} feature in the NIRISS and STIS data proves challenging for the retrieval to fit, and thus removing the species is not detrimental.

\begin{figure*}
    \centering
    \includegraphics[width=\textwidth]{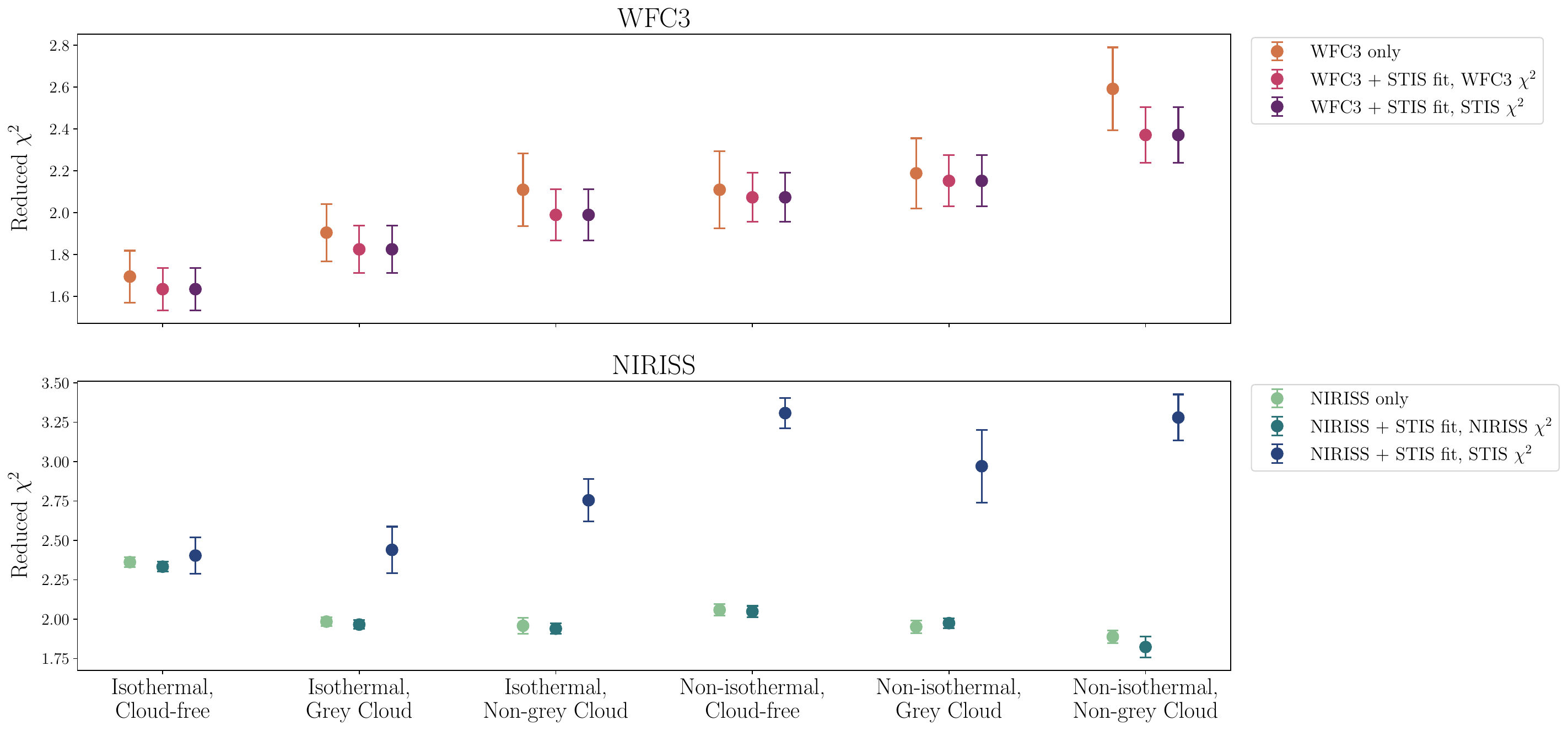}
    \caption{Reduced $\chi^2$ values for retrievals with each model. The top panel shows the reduced $\chi^2$ values for the retrievals on WFC3 data alone, as well as for the joint WFC3 + STIS data. In the joint case, the reduced $\chi^2$ is shown for the WFC3 and STIS sections separately. The bottom panel shows the equivalent for the NIRISS and NIRISS + STIS retrievals.}
    \label{fig:chi_sq}
\end{figure*}

\begin{figure*}
    \centering
    \includegraphics[width=\textwidth]{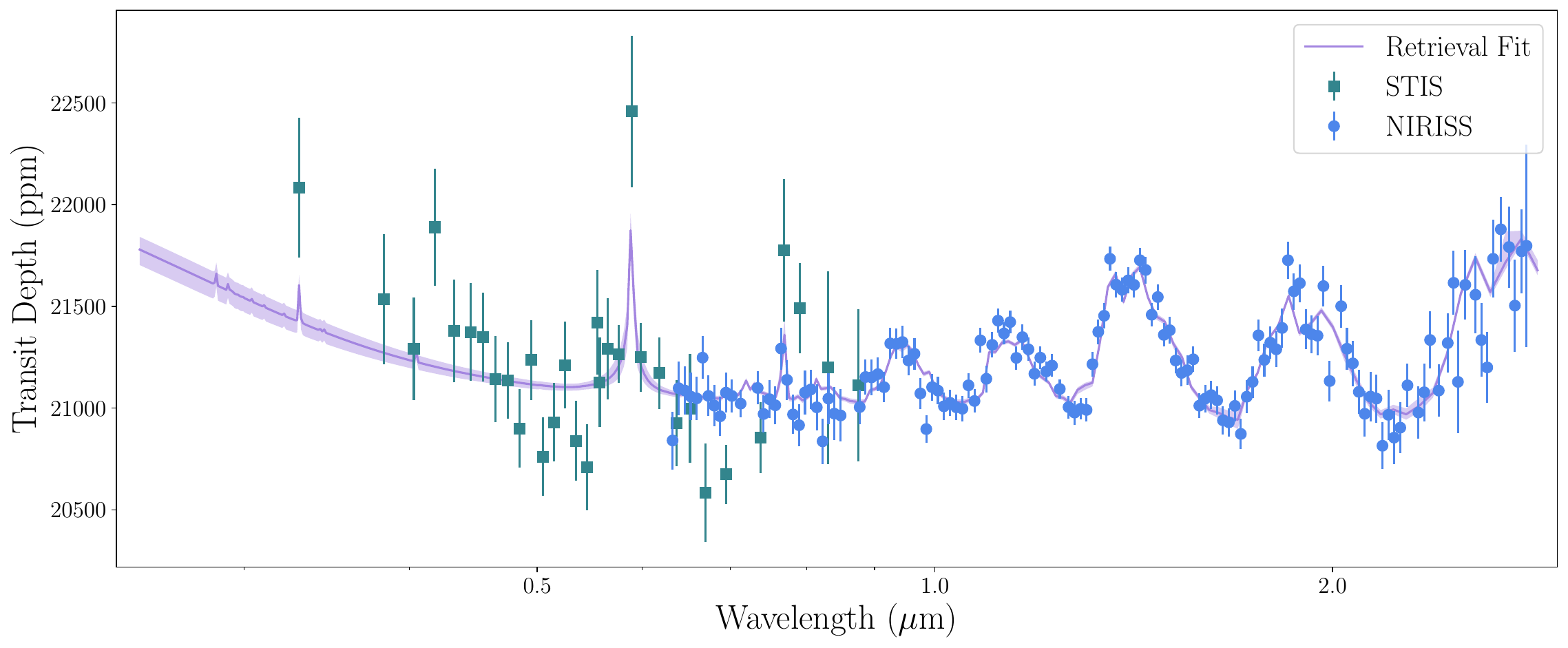}
    \caption{Retrieval fit for NIRISS + STIS using the non-isothermal, non-grey cloud model. The purple line and shaded region show the median and 1-sigma transit depth values.}
    \label{fig:Retrieval_fit_niriss_stis}
\end{figure*}

Figure \ref{fig:chi_sq} shows the reduced $\chi^2$ values for the different retrievals, for the different sections of data. In general, the addition of STIS to the retrieval does not seem to change the reduced $\chi^2$ value for the WFC3 or NIRISS spectra. However, for the NIRISS + STIS retrievals, the reduced $\chi^2$ values for the STIS section are further from 1 in all cases than the NIRISS section. This indicates that, although the addition of STIS data does seem to affect certain model parameters, such as the water abundance, the fit to the data is poor, and thus the effects may not be physical. This is further demonstrated in Figure \ref{fig:Retrieval_fit_niriss_stis}, which shows the posterior fits to the data from the NIRISS + STIS retrieval for the non-isothermal + non-grey cloud model. It is evident that the STIS slope, in particular, is not fit well by this model, despite it having the highest Bayesian evidence. It is unknown whether the poor fit to the STIS data is due to instrument systematics or missing model complexity, such as stellar contamination or a high-altitude haze. Although our non-grey cloud model could fit a thick haze, this would worsen the fit in the infrared region, so is not favoured by the retrieval. In order to fit this very steep optical slope, \cite{fairman24} require an extreme scattering slope of $\gamma < -10$, far beyond the small particle limit of Mie theory (-4) \citep{bohren98}. In turn, this drives the water abundance up to the top of their prior ($\log{X_{\ch{H2O}}}=-1.40^{+0.29}_{-0.76}$), though this value is consistent with our results. However, it is possible to produce a super-Rayleigh slope that could fit very steep optical data with a detached haze layer and a clear atmosphere beneath \citep{barstow17}, or with a photochemical haze formed in a vigorously mixing atmosphere \citep{kawashima19,ohno20}, both of which could provide an explanation for the STIS data. Recent work also considered the possiblity of fitting this data with unocculted active regions in the star, and found the retrieved molecular abundances for the planet were unchanged -- a reassuring result for retrievals on JWST data (Welbanks et al., \textit{in prep}). If the slope is instead due to instrument systematics, this could affect the retrieved molecular abundances. We suggest that one should exercise caution when directly combining STIS data with new JWST transmission spectra, especially when the data was not reduced in a consistent way.

\subsubsection{Shifting STIS}
\label{sec:shifting_stis}

It is a known issue that spectra from different observing instruments can be offset from one another \citep{yip21}. To test this issue, we add a parameter to all our retrievals that shifts the STIS data up or down with respect to the WFC3 or NIRISS data. It turns out that allowing for a shift has little-to-no effect on the retrievals, and in all cases the STIS shift parameter is consistent with zero (see Figure \ref{fig:stis_shift}). This suggests that the issue with fitting the STIS data is not caused by only an offset between the datasets. 

\subsubsection{Removing Sodium}

\begin{figure*}
    \centering
    \includegraphics[width=0.9\textwidth]{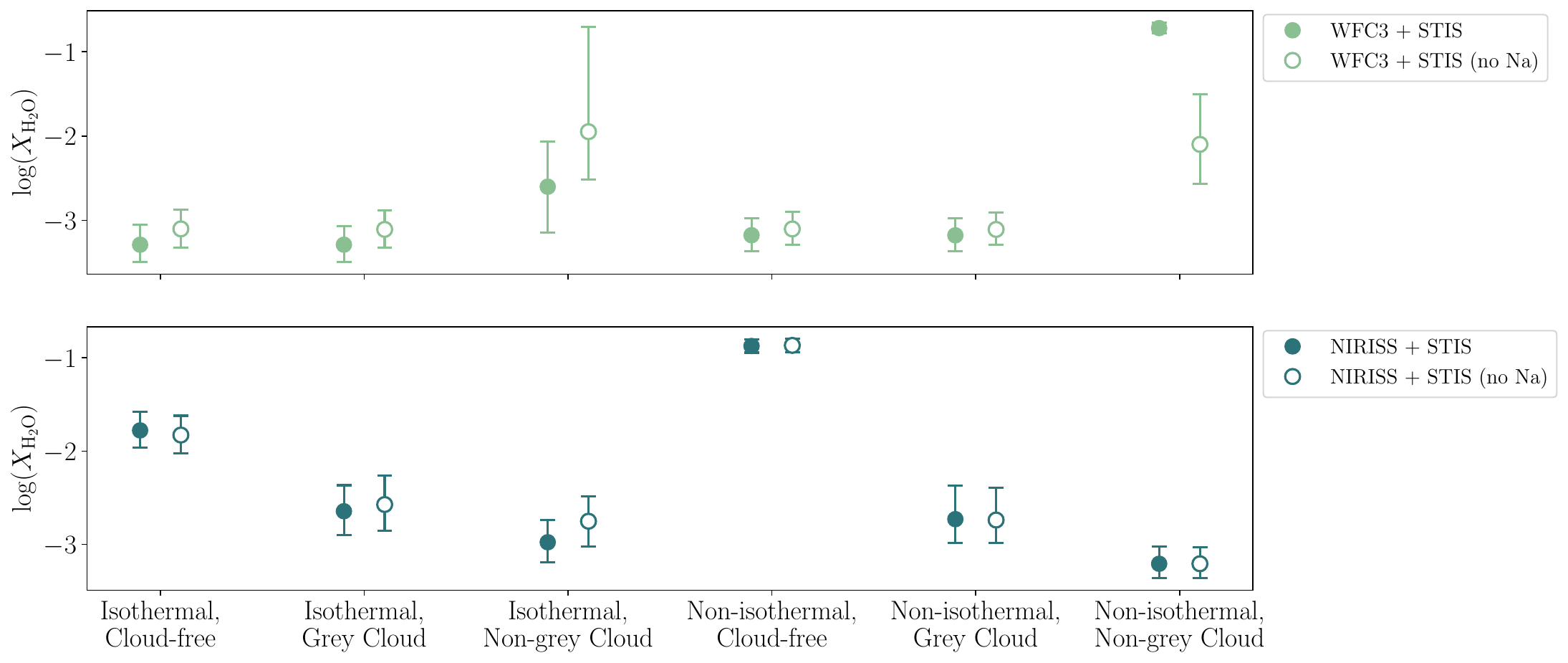}
    \caption{Retrieved water abundances for the full data (filled circles) and for the data with the sodium line removed (empty circles). In the latter case, the sodium abundance and opacity is not included in the retrieval.}
    \label{fig:Retrieval_comparison_noNa}
\end{figure*}

Our combined retrievals demonstrate that the addition of STIS data has a substantial effect on the constraints for water abundance. However, the source of the constraints in retrievals is not always clear, and requires further analysis \citep[e.g.,][]{welbanks23}. The STIS data encompasses a strong and very broad sodium line, as well as an optical slope from clouds/hazes or Rayleigh scattering. In order to test which of these properties is affecting the constraint on water abundance, we perform a set of retrievals on the combined datasets, but remove the STIS data points around the sodium feature. We also remove the sodium opacity from the model, and the sodium abundance is no longer a free parameter in the retrieval. The results for the retrieved water abundances are shown in Figure \ref{fig:Retrieval_comparison_noNa}. In the WFC3 cases, the water abundance remains consistent when sodium is removed, with the exception of the non-grey cloud retrievals. Since these models have more freedom to fit slopes in the optical, it seems that the sodium feature is important for constraining the slope, and thus the spectral continuum. However, in the NIRISS cases, no such effect appears. This is likely due to the larger number of points covering the continuum in the NIRISS data set, meaning the effect of removing sodium is minimised. Overall, this demonstrates that the optical slope is driving the water abundance constraint when STIS data is included, but that flexible cloud models benefit from the additional information from the sodium feature. 

\subsubsection{Alkali Detections}
\label{sec:addingstis_alkalis}

\begin{figure*}
    \centering
    \includegraphics[width=\textwidth]{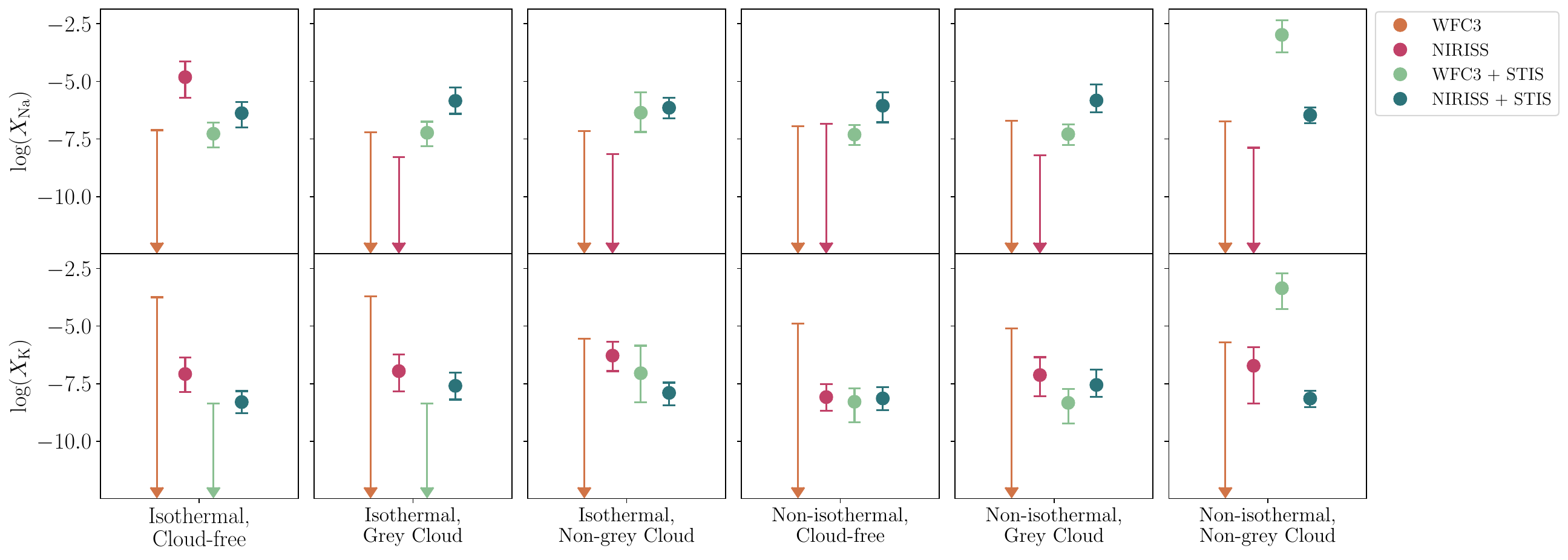}
    \caption{Summary of the retrieval results for the sodium and potassium abundances, for all the models and all the data set combinations. The downward arrows indicate upper limits only.}
    \label{fig:Retrieval_comparison_alkalis}
\end{figure*}

The ERS study of NIRISS data found evidence for \ch{K}, but no evidence for \ch{Na} due to the sodium doublet lying outside the NIRISS wavelength range \citep{feinstein23}. Theoretically, the slope of the \ch{Na} resonance line wings at the blue end of NIRISS could provide some constraints, but it may also be degenerate with other atmospheric properties, such as clouds \citep[e.g.,][]{radica23, taylor23}. The ability to compute refractory-to-volatile ratios could be key for placing constraints on the formation history of planet \citep{lothringer21,schneider21,chachan23}, making detections of these alkalis particularly interesting. Figure \ref{fig:Retrieval_comparison_alkalis} shows the retrieval results for sodium and potassium for the different dataset combinations. In agreement with the ERS study, we find that in order to detect sodium the retrieval requires the STIS wavelengths to cover the sodium doublet. The simplest model, isothermal + cloud-free, does show a detection for the NIRISS-only case, however the Bayesian evidence strongly ruled out this model (Figure \ref{fig:Bayesian_Evidences}). The retrieved \ch{Na} abundance appears fairly consistent across the NIRISS + STIS retrievals, obtaining values of $\sim10^{-6}$. For WFC3 + STIS the abundance of \ch{Na} is fairly consistent across models, with the exception of the non-isothermal + non-grey cloud case, issues of which are explained in Section \ref{sec:addingstis_watertemp}. The abundance is generally slightly lower than the NIRISS + STIS retrievals, probably due to the higher continuum fitting the WFC3 data. When comparing the detection significances for \ch{Na}, as mentioned above, we obtain 3.8$\sigma$ for WFC3 + STIS and 2.3$\sigma$ for NIRISS + STIS. However, note that this is only for the non-isothermal + non-grey cloud retrieval, which obtains substantially higher alkali abundances than other retrievals on WFC3 + STIS.

For potassium, we find the expected result that the NIRISS data alone is able to detect it, though the abundance depends on the model, with values ranging from $\sim10^{-6}$ to $\sim10^{-8}$. The results are mostly consistent when the STIS data is added, showing that the necessary information is already encompassed in the NIRISS data. For WFC3 however, potassium is unsurprisingly not detected, and only when adding STIS and using certain models does one retrieve a constrained abundance for \ch{K}. When it is constrained, the abundance appears to be consistent with the NIRISS retrievals, again with the exception of the non-isothermal + non-grey cloud case. Thus, NIRISS alone can be a powerful tool for diagnosing the \ch{K} content in exoplanetary atmospheres in order to constrain formation histories. Alternatively, one could combine space-based near-infrared data with ground-based optical data in order to constrain the alkali abundances. For the detection significances for \ch{K}, we obtain 2.6$\sigma$ for WFC3 + STIS, and 2.1$\sigma$ for NIRISS. As discussed previously, for NIRISS + STIS the Bayesian evidence favours the retrieval without \ch{K}, likely due to the overlapping but inconsistent \ch{K} feature in the two datasets. This again provides evidence that blindly combining the two datasets could be detrimental to the retrieval results.

\subsection{Cutting NIRISS}
\label{sec:cutting_niriss}

\begin{figure*}
    \centering
    \includegraphics[width=\textwidth]{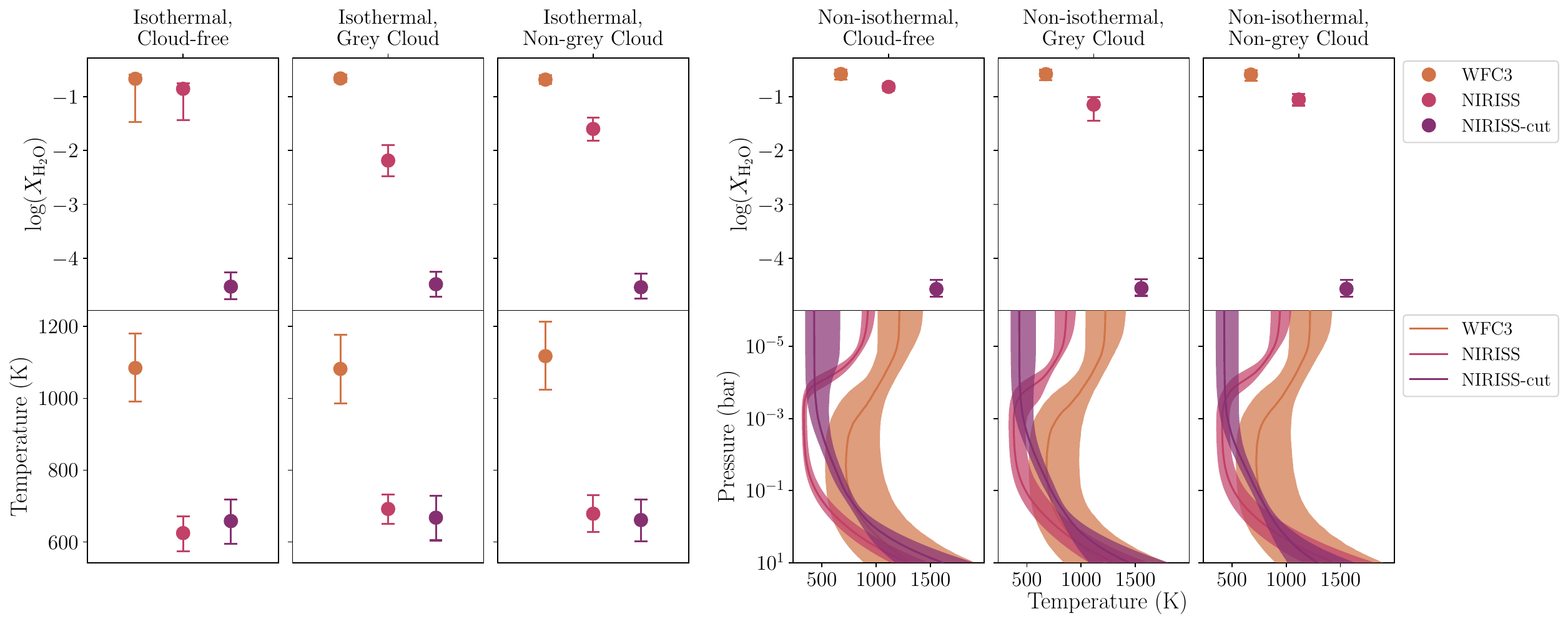}
    \caption{Summary of the retrieval results for all the models, for WFC3, NIRISS, and NIRISS cut to the wavelength range of WFC3.}
    \label{fig:Retrieval_comparison_cutniriss}
\end{figure*}

In this section we want to address the question of whether the biggest gain in information from NIRISS over WFC3 comes from the increased resolution and precision, or the extended wavelength coverage. In order to test this, we perform the retrievals on the NIRISS data cut to the wavelength range of the WFC3 data. Figure \ref{fig:Retrieval_comparison_cutniriss} shows the summary of the results. 

For the water abundance, the NIRISS-cut retrievals obtain substantially lower values than either the WFC3 or full NIRISS retrievals. This is likely due to the lower relative NIRISS points in the \SI{1.4}{\um} water feature, possibly due to different instrument systematics, and the lack of wavelength coverage required to provide the continuum information. As shown in previous studies, and discussed in Section \ref{sec:wfc3_vs_niriss}, the water abundance cannot be constrained in the limited wavelength range of WFC3, except in the cases of very high or very low abundance \citep{line16,welbanks19}. The WFC3 retrieval represents the very high abundance case, where the mean molecular weight is affected, whereas the NIRISS-cut retrieval represents the other extreme, where low water abundance reveals the continuum from CIA at the troughs of the water feature. In reality, the water abundance is likely in-between the two extremes, but this can only be constrained when additional features are encompassed by a wider wavelength range (i.e., the full NIRISS case). This result has serious implications for previous constraints on abundances from WFC3 data, as they are unlikely to be robust, and are potentially very sensitive to the instrument systematics. It also demonstrates how, when the data is scarce, retrievals can struggle to provide good estimates for the errors on their constraints. 

For the isothermal cases, the temperature for the NIRISS-cut retrieval is consistent with the full NIRISS retrieval. This suggests that the temperature is governed by the resolution and precision of the data. For the non-isothermal cases, the picture is more complicated. At the base of the atmosphere, all three profiles agree fairly well, though with quite loose constraints. In the middle of the atmosphere, where the transmission spectrum is probing, the NIRISS-cut profile transitions from agreeing more closely with the WFC3 profile, at pressures around 0.1 bar, to the NIRISS profile, at pressures around $10^{-3}-10^{-4}$ bar. At the top of the atmosphere, the NIRISS-cut profile is substantially colder than either of the others. Interestingly, the NIRISS-cut profiles show no evidence for an inversion, despite this appearing in all the other profiles. These results suggest that in the non-isothermal case, a combination of wavelength coverage and resolution contribute to the temperature profile's constraints. However, as noted before, we do not expect there to be sufficient information in transmission spectra to accurately constrain all levels of the temperature profile for the planet. 

An additional interesting comparison comes from the Bayesian evidences. In the NIRISS-cut retrievals, the models all obtain a Bayes factor of 0.6 or less, giving inconclusive evidence for favouring any of them. This is comparable to the Bayesian evidence from the WFC3 case (upper-left panel of Figure \ref{fig:Bayesian_Evidences}), suggesting that the wavelength coverage is key for distinguishing between different models.

\subsection{Mock Retrievals}
\label{sec:mock_retrievals}

In order to test the conclusions from Sections \ref{sec:wfc3_vs_niriss}, \ref{sec:adding_stis}, and \ref{sec:cutting_niriss}, we perform mock retrievals on simulated data. This removes the differences seen in the real data potentially due to systematics. For each of the six models (isothermal/non-isothermal + cloudfree/grey cloud/non-grey cloud), we simulate the WFC3, NIRISS and STIS data assuming the system parameters of WASP-39 b, and setting the model parameters to the best-fit values from the retrievals on the real NIRISS data. This is with the exception of the water abundance, which is fixed to the same value across all models. This allows us to test each model's ability to retrieve the water abundance accurately, in a consistent way. The error bars are assumed to be the same as for the real WASP-39 b data, though no noise was added to the simulated points. This idealistic case allows us to focus on differences due to only the wavelength coverage and resolution/precision, without being affected by a random noise draw. 

\subsubsection{Full Data}

\begin{figure*}
    \centering
    \includegraphics[width=\textwidth]{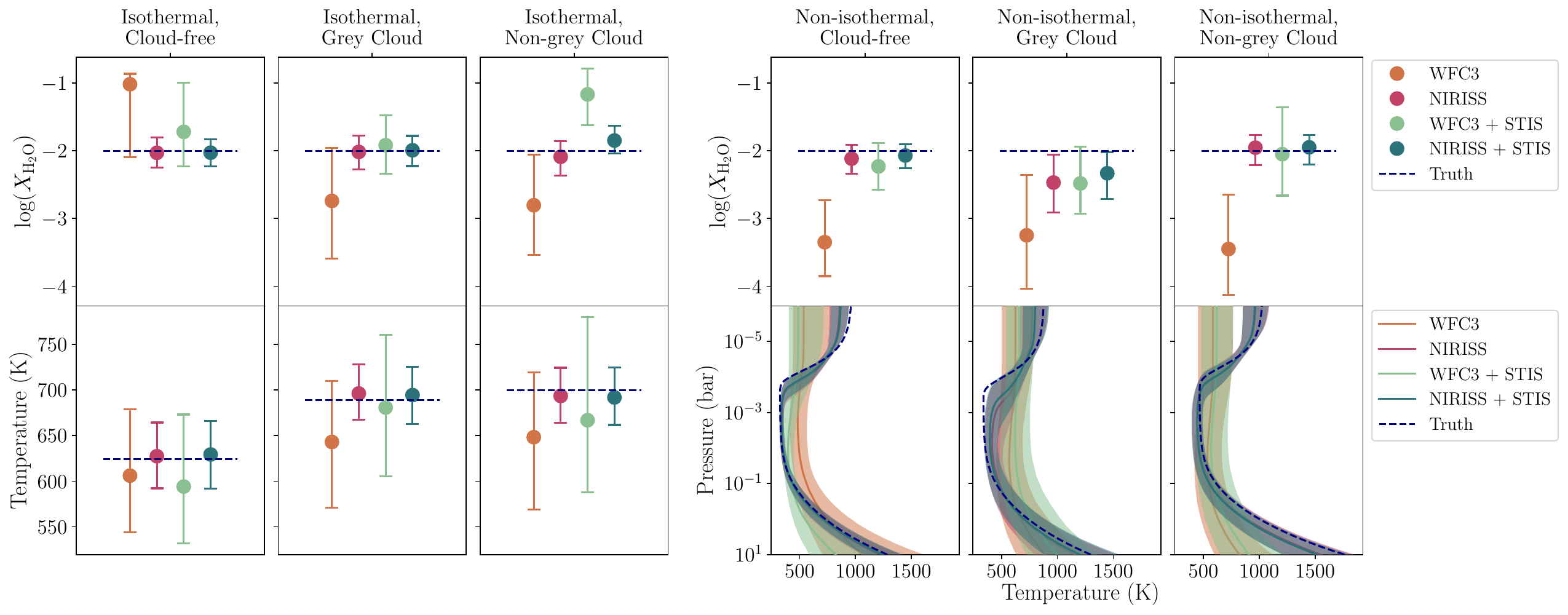}
    \caption{Mock retrieval results for the water abundance and temperature. In this case, the same model was used to generate the mock spectrum and perform the retrieval. The input parameters were set to the best-fit values from the real retrievals on the NIRISS data, and the water abundance was fixed to $10^{-2}$. The true values are shown by the navy dashed lines. The error bars on the mock spectra are assumed to be the same as the real data for WASP-39 b, but no noise was added to the data points. This idealistic case allows us to assess the retrieval's performance in the absolute best-case scenario.}
    \label{fig:mockretrieval_h2o_1e-2}
\end{figure*}

\begin{figure*}
    \centering
    \includegraphics[width=0.8\textwidth]{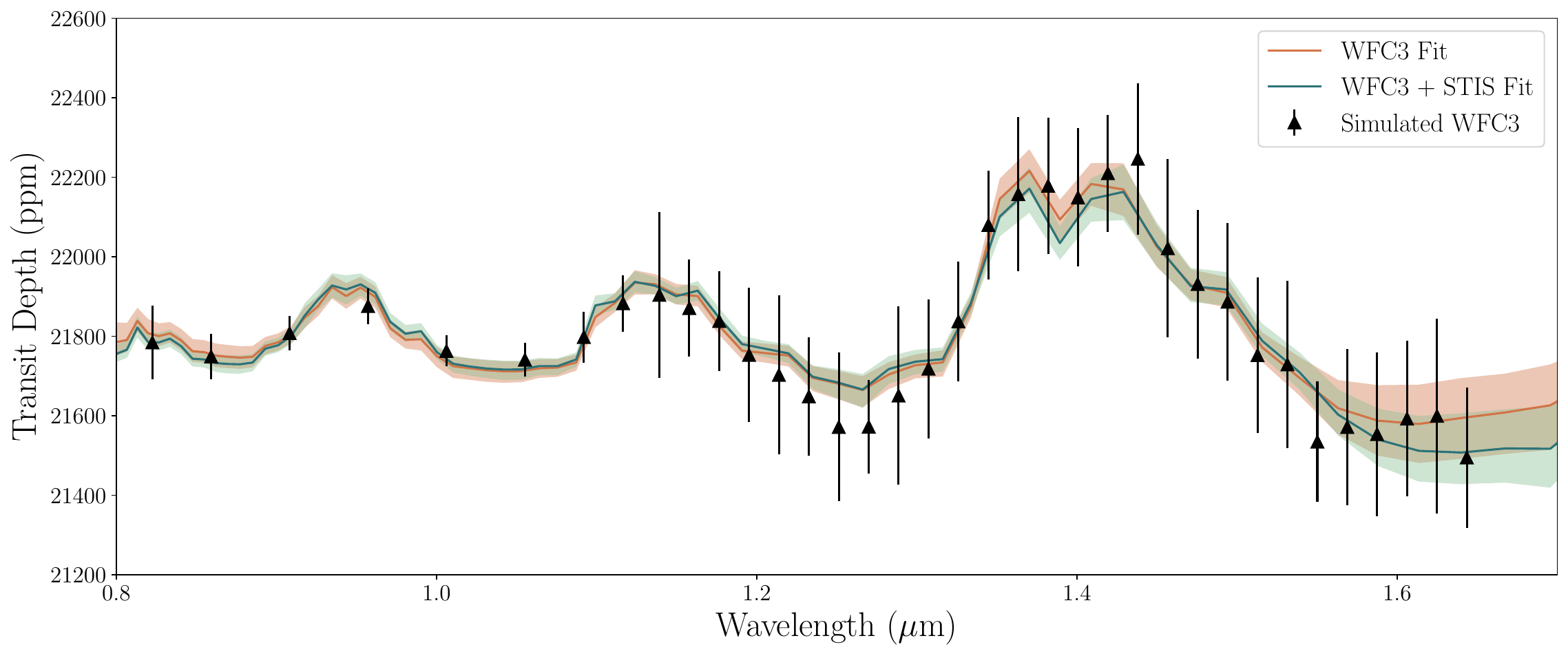}
    \caption{Spectral fits to the WFC3 data for the mock retrievals of the simulated data, assuming a water abundance of $10^{-2}$, using the non-isothermal + non-grey cloud model. The pink line and shaded region show the fit and 1-sigma error from the retrieval on the simulated WFC3 data alone, respectively. The red line and shaded region show the fit and 1-sigma error from the retrieval on the simulated WFC3 and STIS data combined, respectively.}
    \label{fig:spectralfit_wfc3}
\end{figure*}

Figure \ref{fig:mockretrieval_h2o_1e-2} shows the results for the WFC3, NIRISS, WFC3 + STIS, and NIRISS + STIS mock retrievals, for a fixed water abundance of $10^{-2}$. For both water abundance and temperature, the NIRISS and NIRISS + STIS retrievals are able to accurately retrieve the input values across all models. However, for WFC3 and WFC3 + STIS, the water abundance is not always correctly retrieved. In fact, for a number of models, the 1-sigma constraints from the WFC3 retrievals significantly underestimate the water abundance. Adding the STIS data helps correct this in most cases. Figure \ref{fig:spectralfit_wfc3} shows the retrieval fits to the mock WFC3 data for the retrievals on WFC3 and WFC3 + STIS for the non-isothermal + non-grey cloud case. The water features from both fits clearly lie within the WFC3 error bars at \SI{1.4}{\um}, despite corresponding to an abundance difference of 2-3 orders of magnitude. Figure \ref{fig:mock_retrieval_cornerplot} shows the full cornerplot for these retrievals. The true abundances are consistent with the posteriors for WFC3 (as expected due to the lack of noise), but lie in the tail of the distribution for the water abundance. It appears the degeneracy with the cloud top pressure is what leads to the lower water abundances, and only extreme abundances at the high and low end are ruled out (as predicted by \cite{line16}). This experiment also demonstrates how the addition of the STIS wavelengths can improve your retrieval constraints. However, it is the noise from the real STIS data, due to systematics or stellar activity, that can cause problematic biases, as shown in Section \ref{sec:adding_stis}.

These mock retrieval tests shows how the WFC3 retrieval adjusts differently to fit the data, and further emphasises the need for extended wavelength coverage in order to accurately constrain molecular abundances. In the isothermal cases, the temperature is accurately retrieved, though with a looser constraint than for the NIRISS retrievals. In the non-isothermal cases, the profiles appear to overlap with the true profile in the areas probed by transmission spectra (i.e. $10^{-1}-10^{-4}$ bar). However, there are discrepancies in the 1-sigma constraints from the WFC3 and WFC3 + STIS retrievals, likely due to the temperature compensating for degenerate water abundances by adjusting the scale height. 

\subsubsection{Cut NIRISS}

\begin{figure*}
    \centering
    \includegraphics[width=\textwidth]{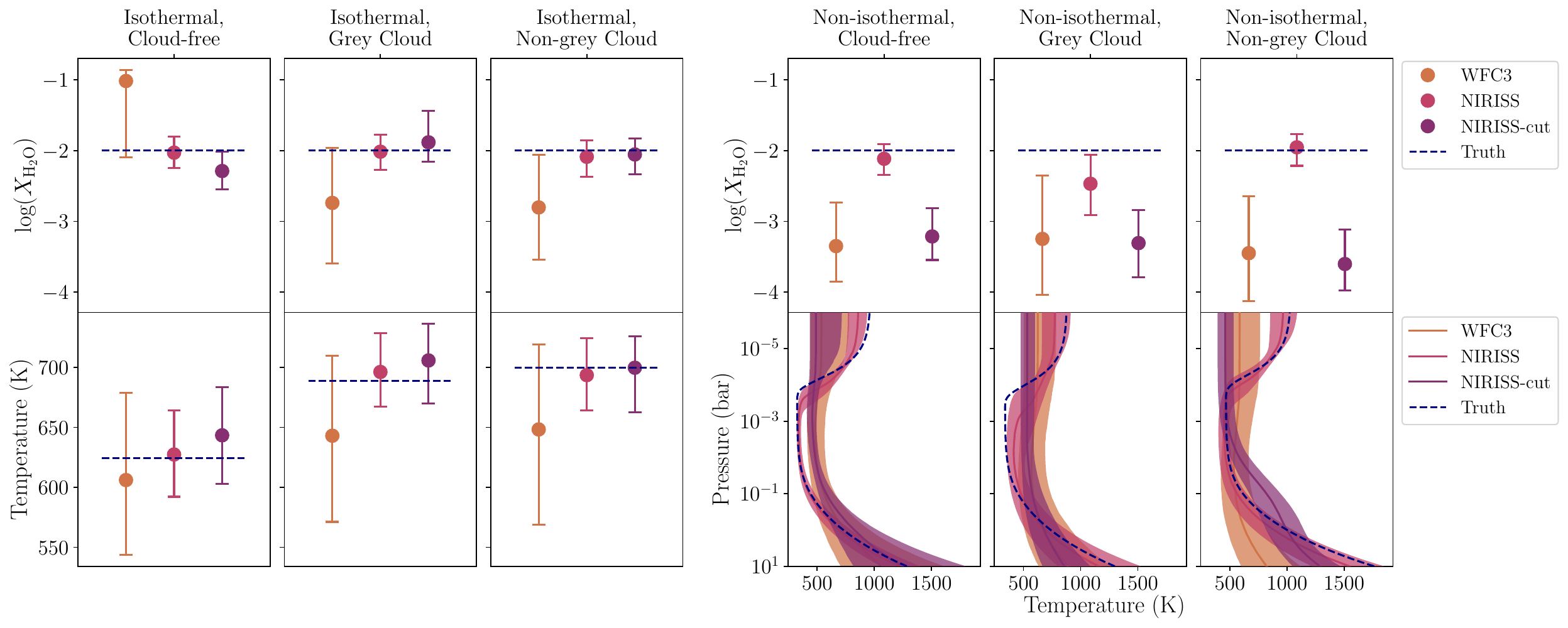}
    \caption{Mock retrieval results for the water abundance and temperature, for the WFC3 and NIRISS simulated data (same as Figure \ref{fig:mockretrieval_h2o_1e-2}), but now including the cut-NIRISS simulated data case as well.}
    \label{fig:mockretrieval_cutniriss}
\end{figure*}

In addition to the mock retrievals on the full data ranges, we also perform a mock retrieval on the NIRISS data cut to the same wavelength range as WFC3, as in Section \ref{sec:cutting_niriss}. Again, the water abundance for the simulated spectrum is fixed to $10^{-2}$, and the other parameters are set to the best-fit values from the retrievals on the real full NIRISS data. Figure \ref{fig:mockretrieval_cutniriss} shows the comparison of the retrieved water abundance and temperature for the WFC3 and NIRISS mock retrievals (same as Figure \ref{fig:mockretrieval_h2o_1e-2}), and the NIRISS-cut mock retrieval. In the isothermal cases, the water abundance from the NIRISS-cut retrievals agrees with the true value, and has a similar constraint to the full NIRISS retrieval. Though the WFC3 retrieval also agrees with the true value here, the constraints are, as expected, much wider due to the larger error bars on the data. However, in the non-isothermal cases, the NIRISS-cut results suffer from the same problem as the WFC3 retrievals, deviating from the true value. Again, the correct water abundance lies in the very edge of the tail of the posterior distribution (not shown). These results suggest that a degeneracy with the temperature profile can also drive the water abundance to different values, even with the higher resolution and smaller error bars associated with NIRISS spectra. As before, the presence of clouds exacerbates the issue. These results provide further evidence in favour of extended wavelength coverage for accurately determining molecular abundances. 

\section{Discussion}
\label{sec:discussion}


The start of the JWST era brings major gains in both resolution and wavelength coverage for exoplanet spectra. This yields a substantial increase in the confidence of our analysis of these atmospheres. It also enables us to probe new molecular features, inaccessible from previous instruments. This helps to build a much more complete picture of these distant planets. However, this increase in detail and precision brings its own issues. The various assumptions made by standard retrieval models may no longer be sufficient to fit this data \citep[e.g.,][]{feng16,macdonald20,taylor20,yang23}, and care must be taken when drawing conclusions from retrieval results. To help mitigate these issues, standard practice has developed to include multiple data reductions and retrievals, using different pipelines and forwards models with distinct origins \citep[e.g.][]{moran23,grant23,welbanks24,yang24}. A thorough exploration of the different approaches to data reduction and forward modelling assumptions in retrieval models can help us fully explore the information content of the JWST era spectra.

With an abundance of archival exoplanet data, and the current operation of HST/STIS and HST/UVIS that access wavelengths not seen by JWST, it is logical to want to combine these data together in order to expand our wavelength coverage and break degeneracies in retrievals. Our mock retrieval results show that, even in the absence of noise, WFC3 data alone is not always able to retrieve the correct water abundance, and the inclusion of STIS wavelengths can help. However, when the real data is used, effects from the instrument's noise can lead to biases in the retrievals when adding STIS data. This highlights some of the issues that can arise when combining datasets from different instruments, especially given they are reduced using different methods with different assumptions. Furthermore, potential variability in the planet can also cause discrepancies in the measurements if they are taken at different times \citep[e.g.][]{meiervaldes23}, leading to issues when combining different epochs of data. It is common to perform retrieval studies, especially population studies, by extracting spectra from the literature instead of homogeneously reducing the data. We have shown that the retrieval results can be biased by one instrument, even when the model lacks the complexity to fit that instrument's data accurately.

A recent study by \cite{lueber24} carried out a detailed analysis of the ERS data of WASP-39b from the different instruments. They found that flexible non-grey cloud models were able to fit spectral features with insufficient wavelength coverage, thus impacting the retrieval results significantly. In the current study, when STIS data is included, the non-grey cloud adjusts to fit the steep slope, affecting the retrieved abundances. However, it's possible that the slope has an alternative source, such as stellar contamination or instrument systematics, and the fit from the non-grey cloud model could be inaccurate. 

As part of the ERS papers, Welbanks et al. (\textit{in prep}) performed a model synthesis study comparing retrieval results from different codes on the full 0.6--12 micron JWST spectra of WASP-39 b. They found a metallicity of 10--50$\times$ solar, depending on modelling assumptions. This is slightly lower than our results, which, for the NIRISS non-isothermal + non-grey cloud retrieval, for example, corresponds to a metallicity of $\sim200$--350$\times$ solar. The wider spectral data likely provides additional constraints on the water vapour abundance beyond that available from NIRISS. Welbanks et al. (\textit{in prep}) note that determining the water abundance of this planet is uniquely challenging, since it occupies a region of parameter space where relatively large changes in the water abundance have a very subtle effect on the spectrum. This is due to the change in mean molecular weight, and therefore scale height, compensating for a change in the absorption due to water.

\section{Conclusion}
\label{sec:conclusion}

In this work, we perform atmospheric retrievals on the JWST/NIRISS and HST/WFC3 spectra of WASP-39 b, to quantify the difference in information content across the two instruments (Section \ref{sec:wfc3_vs_niriss}). We find that even the water abundance, the only molecule confidently probed by the WFC3 wavelength range, can be biased to spurious values due to the lack of continuum information in the short WFC3 window. The NIRISS data provides this essential information, as well as additional absorbers, such as \ch{K}, and clouds. Furthermore, our analysis reveals a substantial gain in the confidence of the water detection for NIRISS over WFC3 ($21.3\sigma$ vs $9.0\sigma$).

We also combine these spectra with STIS data, to expand our comparison (Section \ref{sec:adding_stis}). This additional data affects the retrieved water abundance, and holds further information on clouds and alkali absorption. However, the STIS data also has the power to bias the retrieval results, for NIRISS for example, despite the poor fit to its spectrum. In order to determine the source of the effect on the constraints from STIS, we tried removing the sodium feature from the data. This generally did not affect the retrieved water abundance, suggested that the optical slope from STIS is driving the constraint, as opposed to the sodium line wings. Adding the STIS data to NIRISS also weakens the detection significance of \ch{CO2}, as well as removing the detection of \ch{K} altogether, which shows further evidence for biasing. Additionally, we test if including a shift parameter applied to the STIS data can mitigate some of these issues in the retrievals (Section \ref{sec:shifting_stis}). However, we find this has little effect, and the shift parameter is consistent with zero, indicating that issues with combining these datasets does not stem from offsets alone. Based on our results, we recommend exercising caution when combining future NIRISS and archival STIS exoplanet spectra. 

The ability to measure refractory-to-volatile ratios could be key for constraining planet formation models. We consider the detectability of alkali species (Section \ref{sec:addingstis_alkalis}) from both the NIRISS and STIS spectra, and find that, although \ch{Na} relies on the bluer wavelengths covered by STIS, NIRISS has the ability to constrain \ch{K}. The opportunity to measure the abundances of \ch{K} and \ch{H2O} could motivate a survey of exoplanets using NIRISS to provide a statistical study of K/O ratios, and thus planet formation histories.

We test our retrievals on the NIRISS spectra cut to the wavelength range covered by WFC3, to determine whether each model parameter is affected by the wavelength coverage or the resolution (Section \ref{sec:cutting_niriss}). We find that the water abundance is strongly affected by the wavelength range, as the inclusion of wider wavelengths is essential for setting the continuum. In this limited wavelength range, the water abundance is extremely sensitive to instrument systematics, which has serious implications for previous constraints from WFC3 data. In the isothermal case, the temperature is driven solely by the resolution of the data, with little information gained from wider wavelengths. In the non-isothermal case, the picture is less clear, though these models may be overly complex.

Finally, we perform mock retrievals on simulated WFC3, NIRISS and STIS data. Our results show that, even with no noise added to the data points, the retrieval struggles to accurately constrain the water abundance from WFC3 data alone, further emphasising the need for the wider wavelength coverage provided by the other datasets. These results also provide more robust evidence of our earlier conclusions that the WFC3 retrievals can obtain incorrect results, as only in the mock retrieval case do we know the ground truth.

Our results demonstrate the vast information gained by the advent of JWST, improving our knowledge of the conditions of exoplanet atmospheres. It remains an exciting period for exoplanet characterisation, and the development of more accurate and reliable analysis techniques must be a priority in order to maximise what we can learn from this data. 

\section*{Acknowledgements}

We would like to thank Adina Feinstein for useful comments on the manuscript. 

C.F. acknowledges financial support from the Swiss National Science Foundation (SNSF) Mobility Fellowship under grant no. P500PT\_203110. C.F. and J.L.B. acknowledge financial support from the European Research Council (ERC) under the European Union’s Horizon 2020 research and innovation program under grant agreement no. 805445. J.T.\ acknowledges the support provided by Schmidt Sciences, LLC. M.R. acknowledges financial support from the Natural Sciences and Engineering Research Council of Canada and the Fonds de recherche du Québec --- Nature et technologies. J.K.B. is supported by a Science and Technology Facilities Council Ernest Rutherford Fellowship (ST/T004479/1). G.M. acknowledges fundings from the Severo Ochoa grant CEX2021-001131-S and the Ram\'on y Cajal grant RYC2022-037854-I funded by MCIN/AEI/ 10.13039/501100011033 and FSE+. 
\\\\
\textit{Software:} \textsc{astropy} \citep{astropycollaboration13,astropycollaboration18,astropycollaboration22}, \textsc{numpy} \citep{harris20}, \textsc{scipy} \citep{virtanen20}, \textsc{matplotlib} \citep{hunter07}, \textsc{pymultinest} \citep{buchner14}, \textsc{corner} \citep{foreman-mackey16}, \textsc{seaborn} \citep{waskom21}.

\section*{Data Availability}

The data used in this study is publicly available and published in \cite{sing16}, \cite{wakeford18}, and \cite{feinstein23}. The open-source retrieval codes \texttt{BeAR} and \texttt{CHIMERA} can be found at \url{https://github.com/newstrangeworlds/bear} and \url{https://github.com.mrline/CHIMERA}, respectively. 



\bibliographystyle{mnras}
\bibliography{references} 

\begin{thebibliography}{}
\makeatletter
\relax
\def\mn@urlcharsother{\let\do\@makeother \do\$\do\&\do\#\do\^\do\_\do\%\do\~}
\def\mn@doi{\begingroup\mn@urlcharsother \@ifnextchar [ {\mn@doi@} {\mn@doi@[]}}
\def\mn@doi@[#1]#2{\def\@tempa{#1}\ifx\@tempa\@empty \href {http://dx.doi.org/#2} {doi:#2}\else \href {http://dx.doi.org/#2} {#1}\fi \endgroup}
\def\mn@eprint#1#2{\mn@eprint@#1:#2::\@nil}
\def\mn@eprint@arXiv#1{\href {http://arxiv.org/abs/#1} {{\tt arXiv:#1}}}
\def\mn@eprint@dblp#1{\href {http://dblp.uni-trier.de/rec/bibtex/#1.xml} {dblp:#1}}
\def\mn@eprint@#1:#2:#3:#4\@nil{\def\@tempa {#1}\def\@tempb {#2}\def\@tempc {#3}\ifx \@tempc \@empty \let \@tempc \@tempb \let \@tempb \@tempa \fi \ifx \@tempb \@empty \def\@tempb {arXiv}\fi \@ifundefined {mn@eprint@\@tempb}{\@tempb:\@tempc}{\expandafter \expandafter \csname mn@eprint@\@tempb\endcsname \expandafter{\@tempc}}}

\bibitem[\protect\citeauthoryear{Abel, Frommhold, Li  \& Hunt}{Abel et~al.}{2011}]{abel11}
Abel M.,  Frommhold L.,  Li X.,   Hunt K. L.~C.,  2011, \mn@doi [The Journal of Physical Chemistry A] {10.1021/jp109441f}, 115, 6805

\bibitem[\protect\citeauthoryear{Abel, Frommhold, Li  \& Hunt}{Abel et~al.}{2012}]{abel12}
Abel M.,  Frommhold L.,  Li X.,   Hunt K. L.~C.,  2012, \mn@doi [The Journal of Chemical Physics] {10.1063/1.3676405}, 136, 044319

\bibitem[\protect\citeauthoryear{Ackerman \& Marley}{Ackerman \& Marley}{2001}]{ackerman01}
Ackerman A.~S.,  Marley M.~S.,  2001, \mn@doi [The Astrophysical Journal] {10.1086/321540}, 556, 872

\bibitem[\protect\citeauthoryear{Ahrer et~al.,}{Ahrer et~al.}{2023}]{ahrer23}
Ahrer E.-M.,  et~al., 2023, \mn@doi [Nature] {10.1038/s41586-022-05590-4}, 614, 653

\bibitem[\protect\citeauthoryear{Albert et~al.,}{Albert et~al.}{2023}]{albert23}
Albert L.,  et~al., 2023, \mn@doi [Publications of the Astronomical Society of the Pacific] {10.1088/1538-3873/acd7a3}, 135, 075001

\bibitem[\protect\citeauthoryear{Alderson et~al.,}{Alderson et~al.}{2023}]{alderson23}
Alderson L.,  et~al., 2023, \mn@doi [Nature] {10.1038/s41586-022-05591-3}, 614, 664

\bibitem[\protect\citeauthoryear{Allard, Spiegelman  \& Kielkopf}{Allard et~al.}{2016}]{allard16}
Allard N.~F.,  Spiegelman F.,   Kielkopf J.~F.,  2016, \mn@doi [Astronomy and Astrophysics] {10.1051/0004-6361/201628270}, 589, A21

\bibitem[\protect\citeauthoryear{Allard, Spiegelman, Leininger  \& Molliere}{Allard et~al.}{2019}]{allard19}
Allard N.~F.,  Spiegelman F.,  Leininger T.,   Molliere P.,  2019, \mn@doi [Astronomy and Astrophysics] {10.1051/0004-6361/201935593}, 628, A120

\bibitem[\protect\citeauthoryear{Asplund, Grevesse, Sauval  \& Scott}{Asplund et~al.}{2009}]{asplund09}
Asplund M.,  Grevesse N.,  Sauval A.~J.,   Scott P.,  2009, \mn@doi [Annual Review of Astronomy and Astrophysics] {10.1146/annurev.astro.46.060407.145222}, 47, 481

\bibitem[\protect\citeauthoryear{{Astropy Collaboration} et~al.,}{{Astropy Collaboration} et~al.}{2013}]{astropycollaboration13}
{Astropy Collaboration} et~al., 2013, \mn@doi [Astronomy and Astrophysics] {10.1051/0004-6361/201322068}, 558, A33

\bibitem[\protect\citeauthoryear{{Astropy Collaboration} et~al.,}{{Astropy Collaboration} et~al.}{2018}]{astropycollaboration18}
{Astropy Collaboration} et~al., 2018, \mn@doi [The Astronomical Journal] {10.3847/1538-3881/aabc4f}, 156, 123

\bibitem[\protect\citeauthoryear{{Astropy Collaboration} et~al.,}{{Astropy Collaboration} et~al.}{2022}]{astropycollaboration22}
{Astropy Collaboration} et~al., 2022, \mn@doi [The Astrophysical Journal] {10.3847/1538-4357/ac7c74}, 935, 167

\bibitem[\protect\citeauthoryear{Barstow, Aigrain, Irwin  \& Sing}{Barstow et~al.}{2017}]{barstow17}
Barstow J.~K.,  Aigrain S.,  Irwin P. G.~J.,   Sing D.~K.,  2017, \mn@doi [The Astrophysical Journal] {10.3847/1538-4357/834/1/50}, 834, 50

\bibitem[\protect\citeauthoryear{Bean et~al.,}{Bean et~al.}{2018}]{bean18}
Bean J.~L.,  et~al., 2018, \mn@doi [Publications of the Astronomical Society of the Pacific] {10.1088/1538-3873/aadbf3}, 130, 114402

\bibitem[\protect\citeauthoryear{Bell et~al.,}{Bell et~al.}{2023}]{bell23}
Bell T.~J.,  et~al., 2023, Methane {Throughout} the {Atmosphere} of the {Warm} {Exoplanet} {WASP}-80b, \mn@doi{10.48550/arXiv.2309.04042}, \url {https://ui.adsabs.harvard.edu/abs/2023arXiv230904042B}

\bibitem[\protect\citeauthoryear{Benneke \& Seager}{Benneke \& Seager}{2012}]{benneke12}
Benneke B.,  Seager S.,  2012, \mn@doi [The Astrophysical Journal] {10.1088/0004-637X/753/2/100}, 753, 100

\bibitem[\protect\citeauthoryear{Benneke \& Seager}{Benneke \& Seager}{2013}]{benneke13}
Benneke B.,  Seager S.,  2013, \mn@doi [The Astrophysical Journal] {10.1088/0004-637X/778/2/153}, 778, 153

\bibitem[\protect\citeauthoryear{Benneke et~al.,}{Benneke et~al.}{2024}]{benneke24}
Benneke B.,  et~al., 2024, {JWST} {Reveals} {CH}\$\_4\$, {CO}\$\_2\$, and {H}\$\_2\${O} in a {Metal}-rich {Miscible} {Atmosphere} on a {Two}-{Earth}-{Radius} {Exoplanet}, \mn@doi{10.48550/arXiv.2403.03325}, \url {https://ui.adsabs.harvard.edu/abs/2024arXiv240303325B}

\bibitem[\protect\citeauthoryear{Berta et~al.,}{Berta et~al.}{2012}]{berta12}
Berta Z.~K.,  et~al., 2012, \mn@doi [The Astrophysical Journal] {10.1088/0004-637X/747/1/35}, 747, 35

\bibitem[\protect\citeauthoryear{Blecic, Dobbs-Dixon  \& Greene}{Blecic et~al.}{2017}]{blecic17}
Blecic J.,  Dobbs-Dixon I.,   Greene T.,  2017, \mn@doi [The Astrophysical Journal] {10.3847/1538-4357/aa8171}, 848, 127

\bibitem[\protect\citeauthoryear{Bohren \& Huffman}{Bohren \& Huffman}{1998}]{bohren98}
Bohren C.~F.,  Huffman D.~R.,  1998, Absorption and {Scattering} of {Light} by {Small} {Particles}.
\url {https://ui.adsabs.harvard.edu/abs/1998asls.book.....B}

\bibitem[\protect\citeauthoryear{Buchner et~al.,}{Buchner et~al.}{2014}]{buchner14}
Buchner J.,  et~al., 2014, \mn@doi [Astronomy and Astrophysics] {10.1051/0004-6361/201322971}, 564, A125

\bibitem[\protect\citeauthoryear{Cadieux et~al.,}{Cadieux et~al.}{2024}]{cadieux24}
Cadieux C.,  et~al., 2024, Transmission {Spectroscopy} of the {Habitable} {Zone} {Exoplanet} {LHS} 1140 b with {JWST}/{NIRISS}, \mn@doi{10.48550/arXiv.2406.15136}, \url {https://ui.adsabs.harvard.edu/abs/2024arXiv240615136C}

\bibitem[\protect\citeauthoryear{Carter et~al.,}{Carter et~al.}{2024}]{carter24}
Carter A.~L.,  et~al., 2024, \mn@doi [Nature Astronomy] {10.1038/s41550-024-02292-x}, 8, 1008

\bibitem[\protect\citeauthoryear{Chachan, Knutson, Lothringer  \& Blake}{Chachan et~al.}{2023}]{chachan23}
Chachan Y.,  Knutson H.~A.,  Lothringer J.,   Blake G.~A.,  2023, \mn@doi [The Astrophysical Journal] {10.3847/1538-4357/aca614}, 943, 112

\bibitem[\protect\citeauthoryear{Changeat et~al.,}{Changeat et~al.}{2024}]{changeat24}
Changeat Q.,  et~al., 2024, \mn@doi [The Astrophysical Journal Supplement Series] {10.3847/1538-4365/ad1191}, 270, 34

\bibitem[\protect\citeauthoryear{Constantinou, Madhusudhan  \& Gandhi}{Constantinou et~al.}{2023}]{constantinou23}
Constantinou S.,  Madhusudhan N.,   Gandhi S.,  2023, \mn@doi [The Astrophysical Journal] {10.3847/2041-8213/acaead}, 943, L10

\bibitem[\protect\citeauthoryear{Darveau-Bernier et~al.,}{Darveau-Bernier et~al.}{2022}]{darveau-bernier22}
Darveau-Bernier A.,  et~al., 2022, \mn@doi [Publications of the Astronomical Society of the Pacific] {10.1088/1538-3873/ac8a77}, 134, 094502

\bibitem[\protect\citeauthoryear{Deming et~al.,}{Deming et~al.}{2013}]{deming13}
Deming D.,  et~al., 2013, \mn@doi [The Astrophysical Journal] {10.1088/0004-637X/774/2/95}, 774, 95

\bibitem[\protect\citeauthoryear{Doyon et~al.,}{Doyon et~al.}{2023}]{doyon23}
Doyon R.,  et~al., 2023, \mn@doi [Publications of the Astronomical Society of the Pacific] {10.1088/1538-3873/acd41b}, 135, 098001

\bibitem[\protect\citeauthoryear{Espinoza, Kossakowski  \& Brahm}{Espinoza et~al.}{2019}]{espinoza19}
Espinoza N.,  Kossakowski D.,   Brahm R.,  2019, \mn@doi [Monthly Notices of the Royal Astronomical Society] {10.1093/mnras/stz2688}, 490, 2262

\bibitem[\protect\citeauthoryear{Faedi et~al.,}{Faedi et~al.}{2011}]{faedi11}
Faedi F.,  et~al., 2011, \mn@doi [Astronomy and Astrophysics] {10.1051/0004-6361/201116671}, 531, A40

\bibitem[\protect\citeauthoryear{Fairman, Wakeford  \& MacDonald}{Fairman et~al.}{2024}]{fairman24}
Fairman C.,  Wakeford H.~R.,   MacDonald R.~J.,  2024, The {Importance} of {Optical} {Wavelength} {Data} on {Atmospheric} {Retrievals} of {Exoplanet} {Transmission} {Spectra}, \url {https://ui.adsabs.harvard.edu/abs/2024arXiv240307801F}

\bibitem[\protect\citeauthoryear{Feinstein et~al.,}{Feinstein et~al.}{2023}]{feinstein23}
Feinstein A.~D.,  et~al., 2023, \mn@doi [Nature] {10.1038/s41586-022-05674-1}, 614, 670

\bibitem[\protect\citeauthoryear{Feng, Line, Fortney, Stevenson, Bean, Kreidberg  \& Parmentier}{Feng et~al.}{2016}]{feng16}
Feng Y.~K.,  Line M.~R.,  Fortney J.~J.,  Stevenson K.~B.,  Bean J.,  Kreidberg L.,   Parmentier V.,  2016, \mn@doi [The Astrophysical Journal] {10.3847/0004-637X/829/1/52}, 829, 52

\bibitem[\protect\citeauthoryear{Feroz \& Hobson}{Feroz \& Hobson}{2008}]{feroz08}
Feroz F.,  Hobson M.~P.,  2008, \mn@doi [Monthly Notices of the Royal Astronomical Society] {10.1111/j.1365-2966.2007.12353.x}, 384, 449

\bibitem[\protect\citeauthoryear{Feroz, Hobson  \& Bridges}{Feroz et~al.}{2009}]{feroz09}
Feroz F.,  Hobson M.~P.,   Bridges M.,  2009, \mn@doi [Monthly Notices of the Royal Astronomical Society] {10.1111/j.1365-2966.2009.14548.x}, 398, 1601

\bibitem[\protect\citeauthoryear{Fischer et~al.,}{Fischer et~al.}{2016}]{fischer16}
Fischer P.~D.,  et~al., 2016, \mn@doi [The Astrophysical Journal] {10.3847/0004-637X/827/1/19}, 827, 19

\bibitem[\protect\citeauthoryear{Fisher \& Heng}{Fisher \& Heng}{2018}]{fisher18}
Fisher C.,  Heng K.,  2018, \mn@doi [Monthly Notices of the Royal Astronomical Society] {10.1093/mnras/sty2550}, 481, 4698

\bibitem[\protect\citeauthoryear{Foreman-Mackey}{Foreman-Mackey}{2016}]{foreman-mackey16}
Foreman-Mackey D.,  2016, \mn@doi [Journal of Open Source Software] {10.21105/joss.00024}, 1, 24

\bibitem[\protect\citeauthoryear{Fournier-Tondreau et~al.,}{Fournier-Tondreau et~al.}{2024}]{fournier-tondreau24}
Fournier-Tondreau M.,  et~al., 2024, \mn@doi [Monthly Notices of the Royal Astronomical Society] {10.1093/mnras/stad3813}, 528, 3354

\bibitem[\protect\citeauthoryear{Freedman, Lustig-Yaeger, Fortney, Lupu, Marley  \& Lodders}{Freedman et~al.}{2014}]{freedman14}
Freedman R.~S.,  Lustig-Yaeger J.,  Fortney J.~J.,  Lupu R.~E.,  Marley M.~S.,   Lodders K.,  2014, \mn@doi [The Astrophysical Journal Supplement Series] {10.1088/0067-0049/214/2/25}, 214, 25

\bibitem[\protect\citeauthoryear{Fu et~al.,}{Fu et~al.}{2022}]{fu22}
Fu G.,  et~al., 2022, \mn@doi [The Astrophysical Journal] {10.3847/2041-8213/ac9977}, 940, L35

\bibitem[\protect\citeauthoryear{Gao et~al.,}{Gao et~al.}{2020}]{gao20}
Gao P.,  et~al., 2020, \mn@doi [Nature Astronomy] {10.1038/s41550-020-1114-3}, 4, 951

\bibitem[\protect\citeauthoryear{Gharib-Nezhad, Iyer, Line, Freedman, Marley  \& Batalha}{Gharib-Nezhad et~al.}{2021}]{gharib-nezhad21}
Gharib-Nezhad E.,  Iyer A.~R.,  Line M.~R.,  Freedman R.~S.,  Marley M.~S.,   Batalha N.~E.,  2021, \mn@doi [The Astrophysical Journal Supplement Series] {10.3847/1538-4365/abf504}, 254, 34

\bibitem[\protect\citeauthoryear{Gibson et~al.,}{Gibson et~al.}{2012}]{gibson12}
Gibson N.~P.,  et~al., 2012, \mn@doi [Monthly Notices of the Royal Astronomical Society] {10.1111/j.1365-2966.2012.20655.x}, 422, 753

\bibitem[\protect\citeauthoryear{Grant et~al.,}{Grant et~al.}{2023}]{grant23}
Grant D.,  et~al., 2023, \mn@doi [The Astrophysical Journal] {10.3847/2041-8213/acfc3b}, 956, L32

\bibitem[\protect\citeauthoryear{Griffith}{Griffith}{2014}]{griffith14}
Griffith C.~A.,  2014, \mn@doi [Philosophical Transactions of the Royal Society of London Series A] {10.1098/rsta.2013.0086}, 372, 20130086

\bibitem[\protect\citeauthoryear{Grimm \& Heng}{Grimm \& Heng}{2015}]{grimm15}
Grimm S.~L.,  Heng K.,  2015, \mn@doi [The Astrophysical Journal] {10.1088/0004-637X/808/2/182}, 808, 182

\bibitem[\protect\citeauthoryear{Grimm et~al.,}{Grimm et~al.}{2021}]{grimm21}
Grimm S.~L.,  et~al., 2021, \mn@doi [The Astrophysical Journal Supplement Series] {10.3847/1538-4365/abd773}, 253, 30

\bibitem[\protect\citeauthoryear{Guillot}{Guillot}{2010}]{guillot10}
Guillot T.,  2010, \mn@doi [Astronomy and Astrophysics, Volume 520, id.A27, {\textless}NUMPAGES{\textgreater}13{\textless}/NUMPAGES{\textgreater} pp.] {10.1051/0004-6361/200913396}, 520, A27

\bibitem[\protect\citeauthoryear{Harris et~al.,}{Harris et~al.}{2020}]{harris20}
Harris C.~R.,  et~al., 2020, \mn@doi [Nature] {10.1038/s41586-020-2649-2}, 585, 357

\bibitem[\protect\citeauthoryear{Hunter}{Hunter}{2007}]{hunter07}
Hunter J.~D.,  2007, \mn@doi [Computing in Science \& Engineering] {10.1109/MCSE.2007.55}, 9, 90

\bibitem[\protect\citeauthoryear{{JWST Transiting Exoplanet Community Early Release Science Team} et~al.,}{{JWST Transiting Exoplanet Community Early Release Science Team} et~al.}{2023}]{jwsttransitingexoplanetcommunityearlyreleasescienceteam23}
{JWST Transiting Exoplanet Community Early Release Science Team} et~al., 2023, \mn@doi [Nature] {10.1038/s41586-022-05269-w}, 614, 649

\bibitem[\protect\citeauthoryear{Kawashima \& Ikoma}{Kawashima \& Ikoma}{2019}]{kawashima19}
Kawashima Y.,  Ikoma M.,  2019, \mn@doi [The Astrophysical Journal] {10.3847/1538-4357/ab1b1d}, 877, 109

\bibitem[\protect\citeauthoryear{Kempton et~al.,}{Kempton et~al.}{2023}]{kempton23}
Kempton E. M.~R.,  et~al., 2023, \mn@doi [Nature] {10.1038/s41586-023-06159-5}, 620, 67

\bibitem[\protect\citeauthoryear{Kirk, López-Morales, Wheatley, Weaver, Skillen, Louden, McCormac  \& Espinoza}{Kirk et~al.}{2019}]{kirk19}
Kirk J.,  López-Morales M.,  Wheatley P.~J.,  Weaver I.~C.,  Skillen I.,  Louden T.,  McCormac J.,   Espinoza N.,  2019, \mn@doi [The Astronomical Journal] {10.3847/1538-3881/ab397d}, 158, 144

\bibitem[\protect\citeauthoryear{Kirk et~al.,}{Kirk et~al.}{2024}]{kirk24}
Kirk J.,  et~al., 2024, \mn@doi [The Astronomical Journal] {10.3847/1538-3881/ad19df}, 167, 90

\bibitem[\protect\citeauthoryear{Kitzmann \& Heng}{Kitzmann \& Heng}{2018}]{kitzmann18}
Kitzmann D.,  Heng K.,  2018, \mn@doi [Monthly Notices of the Royal Astronomical Society] {10.1093/mnras/stx3141}, 475, 94

\bibitem[\protect\citeauthoryear{Kitzmann, Heng, Oreshenko, Grimm, Apai, Bowler, Burgasser  \& Marley}{Kitzmann et~al.}{2020}]{kitzmann20}
Kitzmann D.,  Heng K.,  Oreshenko M.,  Grimm S.~L.,  Apai D.,  Bowler B.~P.,  Burgasser A.~J.,   Marley M.~S.,  2020, \mn@doi [The Astrophysical Journal] {10.3847/1538-4357/ab6d71}, 890, 174

\bibitem[\protect\citeauthoryear{Kitzmann, Stock  \& Patzer}{Kitzmann et~al.}{2023}]{kitzmann23}
Kitzmann D.,  Stock J.~W.,   Patzer A. B.~C.,  2023, {FastChem} {Cond}: {Equilibrium} chemistry with condensation and rainout for cool planetary and stellar environments, \mn@doi{10.48550/arXiv.2309.02337}, \url {https://ui.adsabs.harvard.edu/abs/2023arXiv230902337K}

\bibitem[\protect\citeauthoryear{Kramida}{Kramida}{2018}]{kramida18}
Kramida A.,  2018, NIST, 126

\bibitem[\protect\citeauthoryear{Kreidberg et~al.,}{Kreidberg et~al.}{2014a}]{kreidberg14}
Kreidberg L.,  et~al., 2014a, \mn@doi [Nature] {10.1038/nature12888}, 505, 69

\bibitem[\protect\citeauthoryear{Kreidberg et~al.,}{Kreidberg et~al.}{2014b}]{kreidberg14a}
Kreidberg L.,  et~al., 2014b, \mn@doi [The Astrophysical Journal] {10.1088/2041-8205/793/2/L27}, 793, L27

\bibitem[\protect\citeauthoryear{Kurucz \& Bell}{Kurucz \& Bell}{1995}]{kurucz95}
Kurucz R.,  Bell B.,  1995, Robert Kurucz CD-ROM, 23

\bibitem[\protect\citeauthoryear{Lacis \& Oinas}{Lacis \& Oinas}{1991}]{lacis91}
Lacis A.~A.,  Oinas V.,  1991, \mn@doi [Journal of Geophysical Research] {10.1029/90JD01945}, 96, 9027

\bibitem[\protect\citeauthoryear{Lecavelier Des~Etangs, Pont, Vidal-Madjar  \& Sing}{Lecavelier Des~Etangs et~al.}{2008}]{lecavelierdesetangs08}
Lecavelier Des~Etangs A.,  Pont F.,  Vidal-Madjar A.,   Sing D.,  2008, \mn@doi [Astronomy and Astrophysics, Volume 481, Issue 2, 2008, pp.L83-L86] {10.1051/0004-6361:200809388}, 481, L83

\bibitem[\protect\citeauthoryear{Li, Gordon, Rothman, Tan, Hu, Kassi, Campargue  \& Medvedev}{Li et~al.}{2015}]{li15}
Li G.,  Gordon I.~E.,  Rothman L.~S.,  Tan Y.,  Hu S.-M.,  Kassi S.,  Campargue A.,   Medvedev E.~S.,  2015, \mn@doi [The Astrophysical Journal Supplement Series] {10.1088/0067-0049/216/1/15}, 216, 15

\bibitem[\protect\citeauthoryear{Lim et~al.,}{Lim et~al.}{2023}]{lim23}
Lim O.,  et~al., 2023, \mn@doi [The Astrophysical Journal] {10.3847/2041-8213/acf7c4}, 955, L22

\bibitem[\protect\citeauthoryear{Line \& Parmentier}{Line \& Parmentier}{2016}]{line16}
Line M.~R.,  Parmentier V.,  2016, \mn@doi [The Astrophysical Journal] {10.3847/0004-637X/820/1/78}, 820, 78

\bibitem[\protect\citeauthoryear{Line et~al.,}{Line et~al.}{2013}]{line13}
Line M.~R.,  et~al., 2013, \mn@doi [The Astrophysical Journal] {10.1088/0004-637X/775/2/137}, 775, 137

\bibitem[\protect\citeauthoryear{Lothringer, Rustamkulov, Sing, Gibson, Wilson  \& Schlaufman}{Lothringer et~al.}{2021}]{lothringer21}
Lothringer J.~D.,  Rustamkulov Z.,  Sing D.~K.,  Gibson N.~P.,  Wilson J.,   Schlaufman K.~C.,  2021, \mn@doi [The Astrophysical Journal] {10.3847/1538-4357/abf8a9}, 914, 12

\bibitem[\protect\citeauthoryear{Lueber, Kitzmann, Bowler, Burgasser  \& Heng}{Lueber et~al.}{2022}]{lueber22}
Lueber A.,  Kitzmann D.,  Bowler B.~P.,  Burgasser A.~J.,   Heng K.,  2022, \mn@doi [The Astrophysical Journal] {10.3847/1538-4357/ac63b9}, 930, 136

\bibitem[\protect\citeauthoryear{Lueber, Novais, Fisher  \& Heng}{Lueber et~al.}{2024}]{lueber24}
Lueber A.,  Novais A.,  Fisher C.,   Heng K.,  2024, \mn@doi [Astronomy \& Astrophysics] {10.1051/0004-6361/202348802}, 687, A110

\bibitem[\protect\citeauthoryear{Lustig-Yaeger et~al.,}{Lustig-Yaeger et~al.}{2023}]{lustig-yaeger23}
Lustig-Yaeger J.,  et~al., 2023, \mn@doi [Nature Astronomy] {10.1038/s41550-023-02064-z}

\bibitem[\protect\citeauthoryear{MacDonald \& Batalha}{MacDonald \& Batalha}{2023}]{macdonald23}
MacDonald R.~J.,  Batalha N.~E.,  2023, \mn@doi [Research Notes of the American Astronomical Society] {10.3847/2515-5172/acc46a}, 7, 54

\bibitem[\protect\citeauthoryear{MacDonald \& Madhusudhan}{MacDonald \& Madhusudhan}{2017}]{macdonald17b}
MacDonald R.~J.,  Madhusudhan N.,  2017, \mn@doi [The Astrophysical Journal] {10.3847/2041-8213/aa97d4}, 850, L15

\bibitem[\protect\citeauthoryear{MacDonald, Goyal  \& Lewis}{MacDonald et~al.}{2020}]{macdonald20}
MacDonald R.~J.,  Goyal J.~M.,   Lewis N.~K.,  2020, \mn@doi [The Astrophysical Journal] {10.3847/2041-8213/ab8238}, 893, L43

\bibitem[\protect\citeauthoryear{Madhusudhan \& Seager}{Madhusudhan \& Seager}{2009}]{madhusudhan09}
Madhusudhan N.,  Seager S.,  2009, \mn@doi [The Astrophysical Journal] {10.1088/0004-637X/707/1/24}, 707, 24

\bibitem[\protect\citeauthoryear{Madhusudhan, Sarkar, Constantinou, Holmberg, Piette  \& Moses}{Madhusudhan et~al.}{2023}]{madhusudhan23}
Madhusudhan N.,  Sarkar S.,  Constantinou S.,  Holmberg M.,  Piette A. A.~A.,   Moses J.~I.,  2023, \mn@doi [The Astrophysical Journal] {10.3847/2041-8213/acf577}, 956, L13

\bibitem[\protect\citeauthoryear{Meier~Valdés et~al.,}{Meier~Valdés et~al.}{2023}]{meiervaldes23}
Meier~Valdés E.~A.,  et~al., 2023, \mn@doi [Astronomy and Astrophysics] {10.1051/0004-6361/202346050}, 677, A112

\bibitem[\protect\citeauthoryear{Min, Ormel, Chubb, Helling  \& Kawashima}{Min et~al.}{2020}]{min20}
Min M.,  Ormel C.~W.,  Chubb K.,  Helling C.,   Kawashima Y.,  2020, \mn@doi [Astronomy and Astrophysics] {10.1051/0004-6361/201937377}, 642, A28

\bibitem[\protect\citeauthoryear{Moran et~al.,}{Moran et~al.}{2023}]{moran23}
Moran S.~E.,  et~al., 2023, \mn@doi [The Astrophysical Journal] {10.3847/2041-8213/accb9c}, 948, L11

\bibitem[\protect\citeauthoryear{Morello, Changeat, Dyrek, Lagage  \& Tan}{Morello et~al.}{2023}]{morello23}
Morello G.,  Changeat Q.,  Dyrek A.,  Lagage P.~O.,   Tan J.~C.,  2023, \mn@doi [Astronomy and Astrophysics] {10.1051/0004-6361/202346643}, 676, A54

\bibitem[\protect\citeauthoryear{Mugnai, Swain, Estrela  \& Roudier}{Mugnai et~al.}{2024}]{mugnai24}
Mugnai L.~V.,  Swain M.~R.,  Estrela R.,   Roudier G.~M.,  2024, \mn@doi [Monthly Notices of the Royal Astronomical Society] {10.1093/mnras/stae1073}, 531, 35

\bibitem[\protect\citeauthoryear{Nikolov, Sing, Gibson, Fortney, Evans, Barstow, Kataria  \& Wilson}{Nikolov et~al.}{2016}]{nikolov16}
Nikolov N.,  Sing D.~K.,  Gibson N.~P.,  Fortney J.~J.,  Evans T.~M.,  Barstow J.~K.,  Kataria T.,   Wilson P.~A.,  2016, \mn@doi [The Astrophysical Journal] {10.3847/0004-637X/832/2/191}, 832, 191

\bibitem[\protect\citeauthoryear{Ohno \& Kawashima}{Ohno \& Kawashima}{2020}]{ohno20}
Ohno K.,  Kawashima Y.,  2020, \mn@doi [The Astrophysical Journal] {10.3847/2041-8213/ab93d7}, 895, L47

\bibitem[\protect\citeauthoryear{Parmentier \& Guillot}{Parmentier \& Guillot}{2014}]{parmentier14}
Parmentier V.,  Guillot T.,  2014, \mn@doi [Astronomy and Astrophysics] {10.1051/0004-6361/201322342}, 562, A133

\bibitem[\protect\citeauthoryear{Parviainen}{Parviainen}{2023}]{parviainen23}
Parviainen H.,  2023, \mn@doi [Astronomy and Astrophysics] {10.1051/0004-6361/202345937}, 671, L3

\bibitem[\protect\citeauthoryear{Pinhas, Madhusudhan, Gandhi  \& MacDonald}{Pinhas et~al.}{2019}]{pinhas19}
Pinhas A.,  Madhusudhan N.,  Gandhi S.,   MacDonald R.,  2019, \mn@doi [Monthly Notices of the Royal Astronomical Society] {10.1093/mnras/sty2544}, 482, 1485

\bibitem[\protect\citeauthoryear{Polyansky, Kyuberis, Zobov, Tennyson, Yurchenko  \& Lodi}{Polyansky et~al.}{2018}]{polyansky18}
Polyansky O.~L.,  Kyuberis A.~A.,  Zobov N.~F.,  Tennyson J.,  Yurchenko S.~N.,   Lodi L.,  2018, \mn@doi [Monthly Notices of the Royal Astronomical Society] {10.1093/mnras/sty1877}, 480, 2597

\bibitem[\protect\citeauthoryear{Radica}{Radica}{2024}]{radica24a}
Radica M.,  2024, {exoTEDRF}: {An} {EXOplanet} {Transit} and {Eclipse} {Data} {Reduction} {Framework}, \mn@doi{10.48550/arXiv.2407.17541}, \url {https://ui.adsabs.harvard.edu/abs/2024arXiv240717541R}

\bibitem[\protect\citeauthoryear{Radica et~al.,}{Radica et~al.}{2022}]{radica22}
Radica M.,  et~al., 2022, \mn@doi [Publications of the Astronomical Society of the Pacific] {10.1088/1538-3873/ac9430}, 134, 104502

\bibitem[\protect\citeauthoryear{Radica et~al.,}{Radica et~al.}{2023}]{radica23}
Radica M.,  et~al., 2023, \mn@doi [Monthly Notices of the Royal Astronomical Society] {10.1093/mnras/stad1762}, 524, 835

\bibitem[\protect\citeauthoryear{Radica et~al.,}{Radica et~al.}{2024}]{radica24}
Radica M.,  et~al., 2024, \mn@doi [The Astrophysical Journal] {10.3847/2041-8213/ad20e4}, 962, L20

\bibitem[\protect\citeauthoryear{Richard et~al.,}{Richard et~al.}{2012}]{richard12}
Richard C.,  et~al., 2012, \mn@doi [Journal of Quantitative Spectroscopy and Radiative Transfer] {10.1016/j.jqsrt.2011.11.004}, 113, 1276

\bibitem[\protect\citeauthoryear{Rocchetto, Waldmann, Venot, Lagage  \& Tinetti}{Rocchetto et~al.}{2016}]{rocchetto16}
Rocchetto M.,  Waldmann I.~P.,  Venot O.,  Lagage P.~O.,   Tinetti G.,  2016, \mn@doi [The Astrophysical Journal] {10.3847/1538-4357/833/1/120}, 833, 120

\bibitem[\protect\citeauthoryear{Rosich, Herrero, Mallonn, Ribas, Morales, Perger, Anglada-Escudé  \& Granzer}{Rosich et~al.}{2020}]{rosich20}
Rosich A.,  Herrero E.,  Mallonn M.,  Ribas I.,  Morales J.~C.,  Perger M.,  Anglada-Escudé G.,   Granzer T.,  2020, \mn@doi [Astronomy and Astrophysics] {10.1051/0004-6361/202037586}, 641, A82

\bibitem[\protect\citeauthoryear{Rothman et~al.,}{Rothman et~al.}{2010}]{rothman10}
Rothman L.~S.,  et~al., 2010, \mn@doi [Journal of Quantitative Spectroscopy and Radiative Transfer] {10.1016/j.jqsrt.2010.05.001}, 111, 2139

\bibitem[\protect\citeauthoryear{Rustamkulov et~al.,}{Rustamkulov et~al.}{2023}]{rustamkulov23}
Rustamkulov Z.,  et~al., 2023, \mn@doi [Nature] {10.1038/s41586-022-05677-y}, 614, 659

\bibitem[\protect\citeauthoryear{Schlawin et~al.,}{Schlawin et~al.}{2016}]{schlawin16}
Schlawin E.,  et~al., 2016, \mn@doi [Publications of the Astronomical Society of the Pacific] {10.1088/1538-3873/129/971/015001}, 129, 015001

\bibitem[\protect\citeauthoryear{Schneider \& Bitsch}{Schneider \& Bitsch}{2021}]{schneider21}
Schneider A.~D.,  Bitsch B.,  2021, \mn@doi [Astronomy and Astrophysics] {10.1051/0004-6361/202141096}, 654, A72

\bibitem[\protect\citeauthoryear{Sing et~al.,}{Sing et~al.}{2015}]{sing15}
Sing D.~K.,  et~al., 2015, \mn@doi [Monthly Notices of the Royal Astronomical Society] {10.1093/mnras/stu2279}, 446, 2428

\bibitem[\protect\citeauthoryear{Sing et~al.,}{Sing et~al.}{2016}]{sing16}
Sing D.~K.,  et~al., 2016, \mn@doi [Nature] {10.1038/nature16068}, 529, 59

\bibitem[\protect\citeauthoryear{Somogyi \& Yurchenko}{Somogyi \& Yurchenko}{2021}]{somogyi21}
Somogyi W.,  Yurchenko S.~N.,  2021, The {Electric} {Quadrupole} {Spectra} of {Diatomic} {Molecules}, \mn@doi{10.15278/isms.2021.TF03.
}, \url {https://ui.adsabs.harvard.edu/abs/2021isms.confETF03S}

\bibitem[\protect\citeauthoryear{Stevenson et~al.,}{Stevenson et~al.}{2016}]{stevenson16}
Stevenson K.~B.,  et~al., 2016, \mn@doi [Publications of the Astronomical Society of the Pacific] {10.1088/1538-3873/128/967/094401}, 128, 094401

\bibitem[\protect\citeauthoryear{Stock, Kitzmann, Patzer  \& Sedlmayr}{Stock et~al.}{2018}]{stock18}
Stock J.~W.,  Kitzmann D.,  Patzer A. B.~C.,   Sedlmayr E.,  2018, \mn@doi [Monthly Notices of the Royal Astronomical Society] {10.1093/mnras/sty1531}, 479, 865

\bibitem[\protect\citeauthoryear{Stock, Kitzmann  \& Patzer}{Stock et~al.}{2022}]{stock22}
Stock J.~W.,  Kitzmann D.,   Patzer A. B.~C.,  2022, \mn@doi [Monthly Notices of the Royal Astronomical Society] {10.1093/mnras/stac2623}, 517, 4070

\bibitem[\protect\citeauthoryear{Taylor, Parmentier, Irwin, Aigrain, Lee  \& Krissansen-Totton}{Taylor et~al.}{2020}]{taylor20}
Taylor J.,  Parmentier V.,  Irwin P. G.~J.,  Aigrain S.,  Lee E. K.~H.,   Krissansen-Totton J.,  2020, \mn@doi [Monthly Notices of the Royal Astronomical Society] {10.1093/mnras/staa552}, 493, 4342

\bibitem[\protect\citeauthoryear{Taylor et~al.,}{Taylor et~al.}{2023}]{taylor23}
Taylor J.,  et~al., 2023, \mn@doi [Monthly Notices of the Royal Astronomical Society] {10.1093/mnras/stad1547}, 524, 817

\bibitem[\protect\citeauthoryear{Trotta}{Trotta}{2008}]{trotta08}
Trotta R.,  2008, \mn@doi [Contemporary Physics] {10.1080/00107510802066753}, 49, 71

\bibitem[\protect\citeauthoryear{Tsai et~al.,}{Tsai et~al.}{2023}]{tsai23}
Tsai S.-M.,  et~al., 2023, \mn@doi [Nature] {10.1038/s41586-023-05902-2}, 617, 483

\bibitem[\protect\citeauthoryear{Tsiaras et~al.,}{Tsiaras et~al.}{2018}]{tsiaras18}
Tsiaras A.,  et~al., 2018, \mn@doi [The Astronomical Journal] {10.3847/1538-3881/aaaf75}, 155, 156

\bibitem[\protect\citeauthoryear{Vardya}{Vardya}{1962}]{vardya62}
Vardya M.~S.,  1962, \mn@doi [The Astrophysical Journal] {10.1086/147269}, 135, 303

\bibitem[\protect\citeauthoryear{Virtanen et~al.,}{Virtanen et~al.}{2020}]{virtanen20}
Virtanen P.,  et~al., 2020, \mn@doi [Nature Methods] {10.1038/s41592-019-0686-2}, 17, 261

\bibitem[\protect\citeauthoryear{Wakeford et~al.,}{Wakeford et~al.}{2013}]{wakeford13}
Wakeford H.~R.,  et~al., 2013, \mn@doi [Monthly Notices of the Royal Astronomical Society] {10.1093/mnras/stt1536}, 435, 3481

\bibitem[\protect\citeauthoryear{Wakeford et~al.,}{Wakeford et~al.}{2018}]{wakeford18}
Wakeford H.~R.,  et~al., 2018, \mn@doi [The Astronomical Journal] {10.3847/1538-3881/aa9e4e}, 155, 29

\bibitem[\protect\citeauthoryear{Waskom}{Waskom}{2021}]{waskom21}
Waskom M.~L.,  2021, \mn@doi [Journal of Open Source Software] {10.21105/joss.03021}, 6, 3021

\bibitem[\protect\citeauthoryear{Welbanks, Madhusudhan, Allard, Hubeny, Spiegelman  \& Leininger}{Welbanks et~al.}{2019}]{welbanks19}
Welbanks L.,  Madhusudhan N.,  Allard N.~F.,  Hubeny I.,  Spiegelman F.,   Leininger T.,  2019, \mn@doi [The Astrophysical Journal] {10.3847/2041-8213/ab5a89}, 887, L20

\bibitem[\protect\citeauthoryear{Welbanks, McGill, Line  \& Madhusudhan}{Welbanks et~al.}{2023}]{welbanks23}
Welbanks L.,  McGill P.,  Line M.,   Madhusudhan N.,  2023, \mn@doi [The Astronomical Journal] {10.3847/1538-3881/acab67}, 165, 112

\bibitem[\protect\citeauthoryear{Welbanks et~al.,}{Welbanks et~al.}{2024}]{welbanks24}
Welbanks L.,  et~al., 2024, \mn@doi [Nature] {10.1038/s41586-024-07514-w}, 630, 836

\bibitem[\protect\citeauthoryear{Yang, Irwin  \& Barstow}{Yang et~al.}{2023}]{yang23}
Yang J.,  Irwin P. G.~J.,   Barstow J.~K.,  2023, \mn@doi [Monthly Notices of the Royal Astronomical Society] {10.1093/mnras/stad2555}, 525, 5146

\bibitem[\protect\citeauthoryear{Yang et~al.,}{Yang et~al.}{2024}]{yang24}
Yang J.,  et~al., 2024, \mn@doi [Monthly Notices of the Royal Astronomical Society] {10.1093/mnras/stae1427}, 532, 460

\bibitem[\protect\citeauthoryear{Yip, Changeat, Edwards, Morvan, Chubb, Tsiaras, Waldmann  \& Tinetti}{Yip et~al.}{2021}]{yip21}
Yip K.~H.,  Changeat Q.,  Edwards B.,  Morvan M.,  Chubb K.~L.,  Tsiaras A.,  Waldmann I.~P.,   Tinetti G.,  2021, \mn@doi [The Astronomical Journal] {10.3847/1538-3881/abc179}, 161, 4

\bibitem[\protect\citeauthoryear{Yurchenko \& Tennyson}{Yurchenko \& Tennyson}{2014}]{yurchenko14}
Yurchenko S.~N.,  Tennyson J.,  2014, \mn@doi [Monthly Notices of the Royal Astronomical Society] {10.1093/mnras/stu326}, 440, 1649

\bibitem[\protect\citeauthoryear{Yurchenko, Tennyson, Barber  \& Thiel}{Yurchenko et~al.}{2013}]{yurchenko13}
Yurchenko S.~N.,  Tennyson J.,  Barber R.~J.,   Thiel W.,  2013, \mn@doi [Journal of Molecular Spectroscopy] {10.1016/j.jms.2013.05.014}, 291, 69

\makeatother
\end{thebibliography}



\appendix

\section{Additional Plots}

\begin{figure*}
    \centering
    \includegraphics[width=\textwidth]{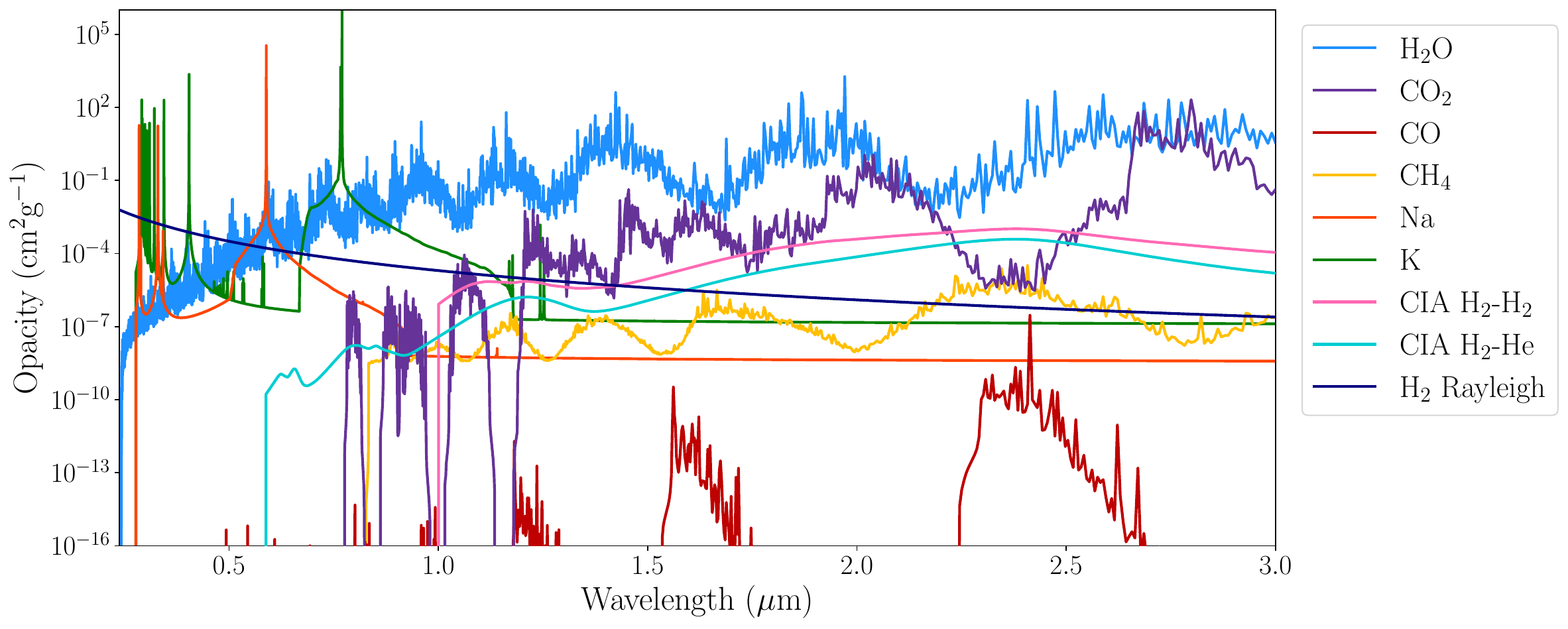}
    \caption{Opacities as a function of wavelength for the opacity sources in the current study, computed at a temperature of 1000 K and a pressure of 0.01 bar. The opacities are weighted by their mass mixing ratios corresponding to the best-fit model from the non-isothermal + non-grey cloud retrieval on the NIRISS data. They are shown over the wavelength range covered by the three instruments in this study -- NIRISS, WFC3 and STIS.}
    \label{fig:opacities}
\end{figure*}

Figure \ref{fig:opacities} shows the opacities of all the species considered in this study, over the wavelength range covered by NIRISS, WFC3 and STIS. Some clear features can be matched to the data in Figure \ref{fig:wasp39b_data} by-eye, such as the sodium and potassium lines, as well as the water absorption features. 

\begin{figure*}
    \centering
    \includegraphics[width=\textwidth]{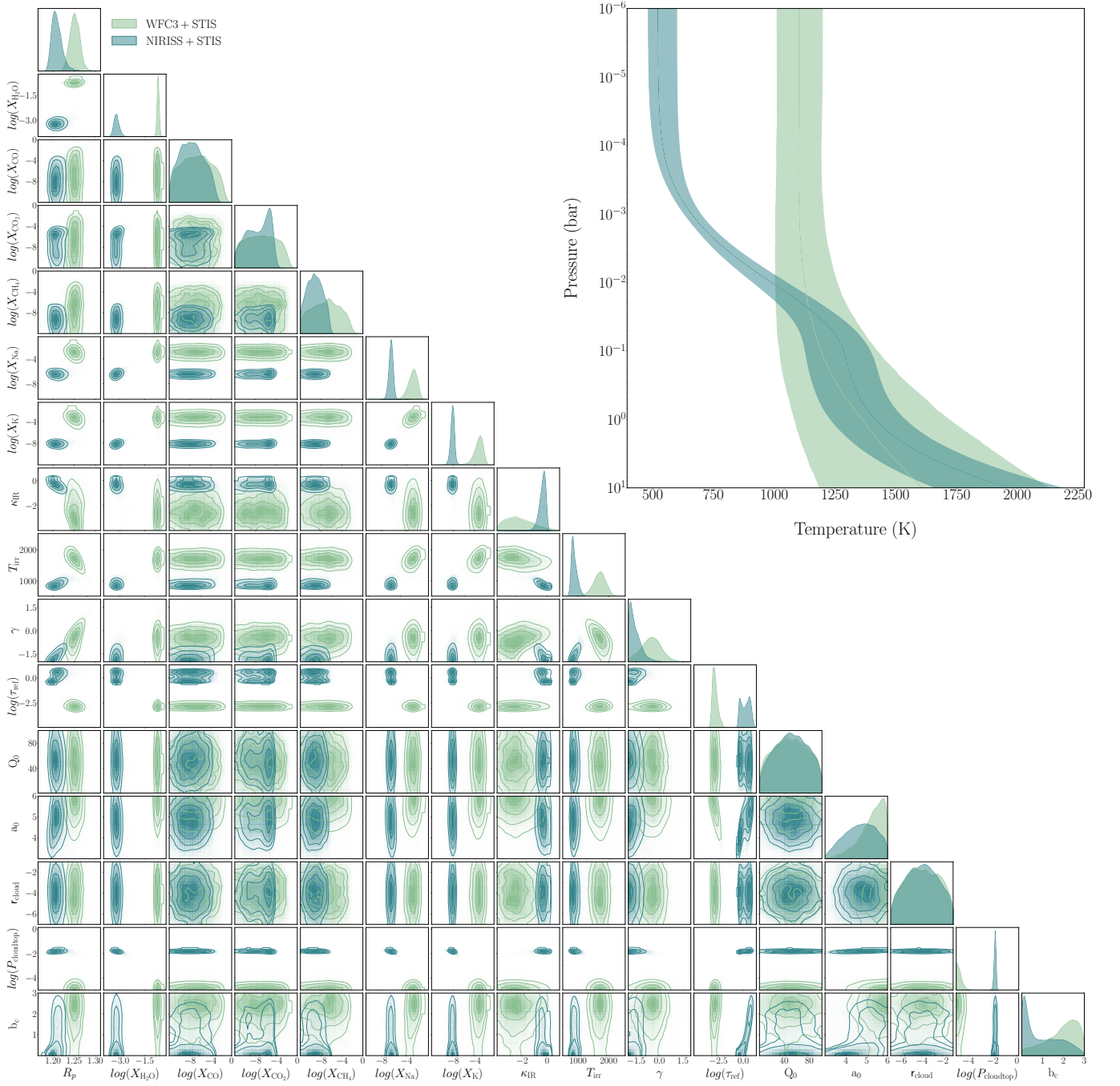}
    \caption{Retrievals using the non-isothermal + non-grey cloud model for the WFC3 + STIS (light green) and the NIRISS + STIS (dark green) data. The median and 1-sigma limits of the retrieved temperature-pressure profiles are shown in the top right corner.}
    \label{fig:WFC3_NIRISS_STIS_retrieval}
\end{figure*}

Figure \ref{fig:WFC3_NIRISS_STIS_retrieval} shows the retrieval results for the non-isothermal + non-grey cloud model applied to the combination of WFC3 + STIS data, and the NIRISS + STIS data. The addition of STIS to WFC3 is causing some of the cloud parameters to be forced to extreme values, hitting the edges of their priors. 

\begin{figure*}
    \centering
    \includegraphics[width=0.9\textwidth]{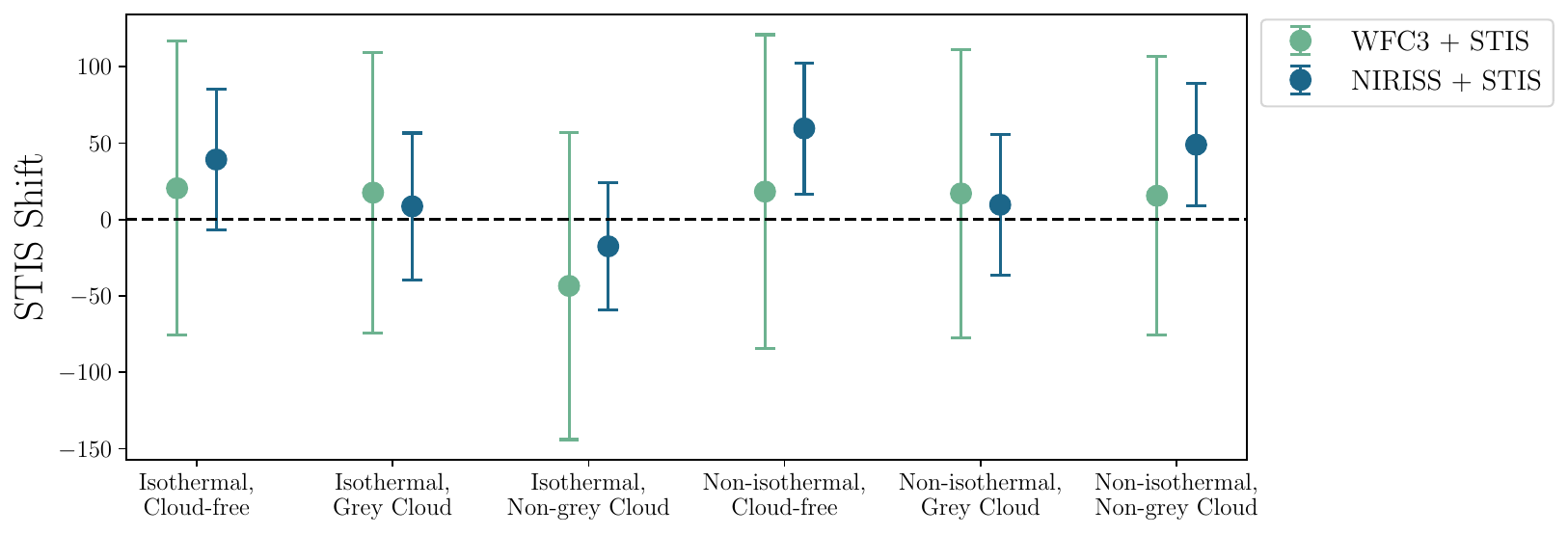}
    \caption{Median and 1-sigma error bars on a shift parameter applied to the STIS data, for the retrievals combined with WFC3 and NIRISS for all six models.}
    \label{fig:stis_shift}
\end{figure*}

Figure \ref{fig:stis_shift} shows the retrieved median values and 1-sigma error bars of the shift parameter applied to the STIS data for all the different models when combined with the WFC3 and NIRISS data. Adding this parameter does not affect the retrieval results, and the consistency with zero demonstrates that there is no evidence for an absolute shift between the datasets. 

\begin{figure*}
    \centering
    \includegraphics[width=\textwidth]{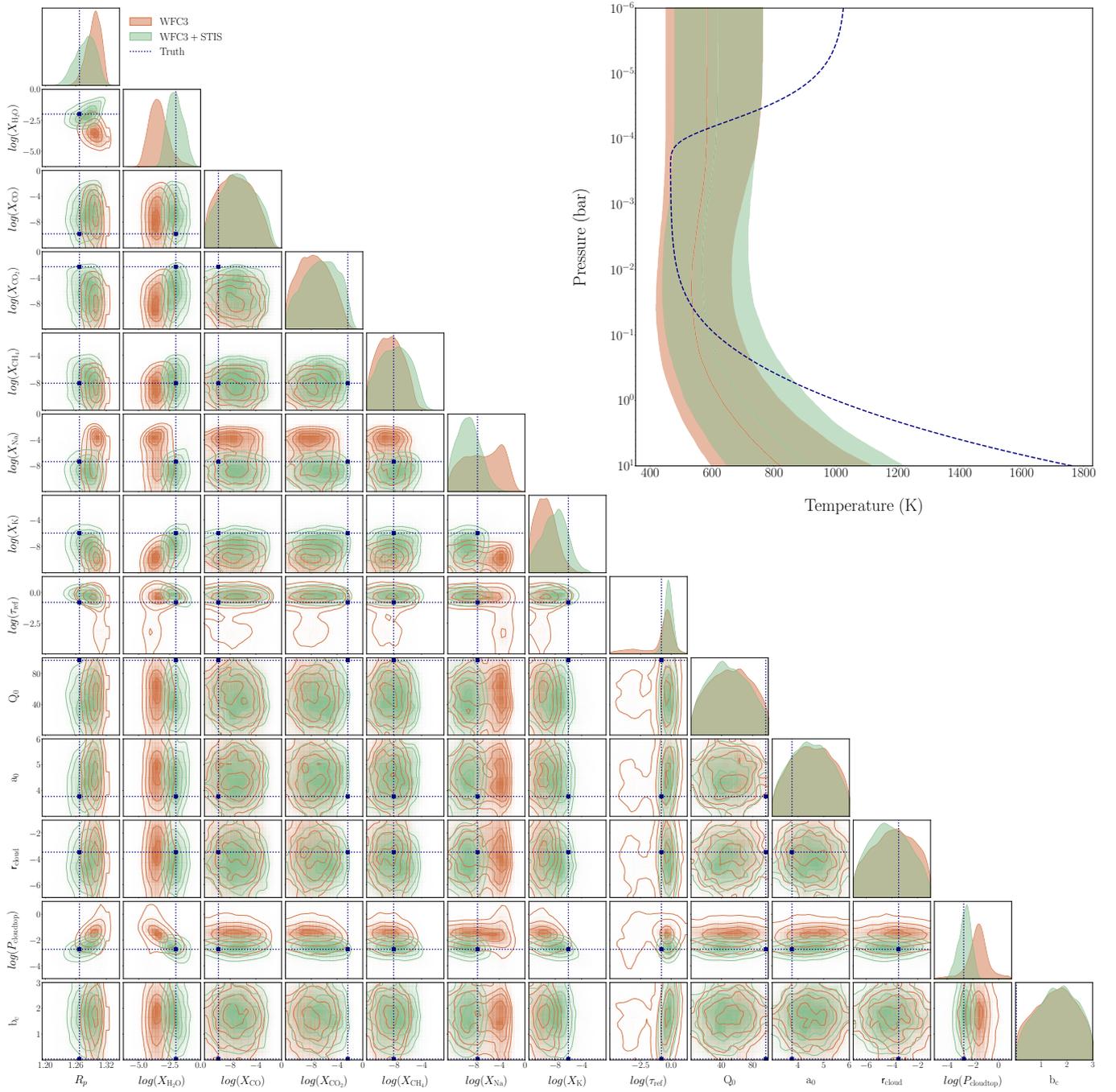}
    \caption{Mock retrieval on the WFC3 (purple) and WFC3 + STIS (green) simulated data, assuming the non-isothermal + non-greycloud model. The navy dashed lines show the true values used for the simulation.}
    \label{fig:mock_retrieval_cornerplot}
\end{figure*}

\section{Retrieved values}

Tables \ref{tab:retrieved_params} and \ref{tab:retrieved_cloud_params} show the retrieved values for each of the datasets, for their best-fit models. 

\begin{table*}
    \centering
    \begin{tabular}{cccccccccc}
         Dataset&  Best Model &  $R_{\rm P}$ ($R_{\rm J}$)&  T (K)&  $\log(X_{\rm H_2O})$&  $\log(X_{\rm CO})$&  $\log(X_{\rm CO_2})$&  $\log(X_{\rm CH_4})$&  $\log(X_{\rm Na})$& $\log(X_{\rm K})$ \\
         \hline
         WFC3& \begin{tabular}{@{}c@{}}Non-isothermal \\ Non-grey Cloud\end{tabular} & $1.28^{+0.01}_{-0.02}$ & $808^{+241}_{-219}$ & $-0.59^{+0.08}_{-0.12}$ & $-6.86^{+3.16}_{-3.01}$ & $-6.83^{+3.1}_{-3.05}$ & $-7.06^{+2.99}_{-2.93}$ & $-6.75^{+3.08}_{-3.02}$ & $-5.71^{+2.55}_{-3.35}$ \\
         \\
         WFC3 + STIS&  \begin{tabular}{@{}c@{}}Isothermal \\ Non-grey Cloud\end{tabular} & $1.28\pm0.01$ & $1126^{+99}_{-97}$ & $-2.6^{+0.53}_{-0.54}$ & $-0.84^{+0.07}_{-0.08}$ & $-7.54^{+2.92}_{-2.95}$ & $-8.22^{+2.5}_{-2.45}$ & $-6.36^{+0.89}_{-0.84}$ & $-7.05^{+1.19}_{-1.25}$ \\
         \\
         NIRISS&  \begin{tabular}{@{}c@{}}Non-isothermal \\ Non-grey Cloud\end{tabular} & $1.28\pm0.01$ & $409^{+69}_{-53}$ & $-1.05^{+0.1}_{-0.11}$ & $-7.01^{+2.99}_{-3.12}$ & $-2.72^{+0.31}_{-0.51}$ & $-8.74^{+2.1}_{-2.05}$ & $-7.88^{+2.56}_{-2.61}$ & $-6.72^{+0.79}_{-1.64}$ \\
         \\
         NIRISS + STIS&  \begin{tabular}{@{}c@{}}Non-isothermal \\ Non-grey Cloud\end{tabular} & 
         $1.21\pm0.01$ & $658^{+37}_{-32}$ & $-3.21^{+0.18}_{-0.16}$ & $-7.77^{+2.64}_{-2.42}$ & $-7.33^{+2.11}_{-2.72}$ & $-9.21^{+1.63}_{-1.6}$ & $-6.46^{+0.33}_{-0.36}$ & $-8.14^{+0.33}_{-0.36}$ \\
    \end{tabular}
    \caption{Retrieved radius, temperature, and molecular abundances for the best model (favoured by the Bayesian Evidence) for each data set. For temperature, if the best model is non-isothermal, the quoted temperature is the value at the 1mbar pressure level.}
    \label{tab:retrieved_params}
\end{table*}

\begin{table*}
    \centering
    \begin{tabular}{cccccccc}
        Dataset & Best Model & $\log(\tau_{\rm cloud})$& $Q_0$& $a_0$& $r_{\rm cloud}$&$\log(P_{\rm cloudtop})$ &$b_c$ \\
        \hline
        WFC3 & \begin{tabular}{@{}c@{}}Non-isothermal \\ Non-grey Cloud\end{tabular} & $-1.88^{+1.61}_{-1.79}$ & $54.25^{+26.97}_{-29.21}$ & $4.4^{+0.91}_{-0.81}$ & $-4.01\pm1.75$ & $-1.9^{+1.68}_{-1.41}$ & $1.6^{+0.8}_{-0.88}$ \\
        WFC3 + STIS & \begin{tabular}{@{}c@{}}Isothermal \\ Non-grey Cloud\end{tabular} & $-3.31^{+0.5}_{-0.43}$ & $51.02^{+32.01}_{-32.92}$ & $5.02^{+0.68}_{-1.07}$ & $-4.03^{+1.94}_{-1.93}$ & $-4.26^{+1.36}_{-0.53}$ & $1.4^{+1.03}_{-0.93}$ \\
        NIRISS & \begin{tabular}{@{}c@{}}Non-isothermal \\ Non-grey Cloud\end{tabular} & $-0.18^{+0.32}_{-0.31}$ & $49.82^{+31.34}_{-30.98}$ & $4.89^{+0.75}_{-0.98}$ & $-4.03^{+1.86}_{-1.88}$ & $-2.45\pm0.24$ & $1.58^{+0.9}_{-1.03}$\\
        NIRISS + STIS & \begin{tabular}{@{}c@{}}Non-isothermal \\ Non-grey Cloud\end{tabular} & $0.24^{+0.45}_{-0.56}$ & $51.46^{+28.4}_{-28.37}$ & $4.71^{+0.75}_{-0.85}$ & $-4.01^{+1.69}_{-1.68}$ & $-1.81^{+0.08}_{-0.1}$ & $0.73^{+1.3}_{-0.63}$ \\
    \end{tabular}
    \caption{Retrieved cloud parameters for the best model (favoured by the Bayesian Evidence) for each data set.}
    \label{tab:retrieved_cloud_params}
\end{table*}

\bsp	
\label{lastpage}
\end{document}